\documentclass[%
 reprint,
 superscriptaddress,
 nofootinbib,
 amsmath,amssymb,
 aps,
 prd,
]{revtex4-1}

\usepackage{graphicx}
\usepackage{dcolumn}
\usepackage{bm}
\usepackage{xcolor, ulem}
\usepackage[colorlinks]{hyperref}
\usepackage{hyperref}
\usepackage{times}
\usepackage[mathlines]{lineno}
\usepackage{subfigure}
\def\mnras{Mon. Not. Roy. Astron. Soc. }%
\renewcommand{\d}{\mathrm{d}}
\definecolor{darkgreen}{rgb}{0.0, 0.2, 0.13}

\begin{document}

\hfill{IPPP/19/59}

\title{High-redshift test of gravity using enhanced growth of small structures\\ probed by the neutral hydrogen distribution}
\author{Matteo Leo}
\email{matteo.leo@durham.ac.uk}
 \affiliation{Institute for Particle Physics Phenomenology, Department of Physics, Durham University, Durham DH1 3LE, U.K.}
\affiliation{
 Institute for Computational Cosmology, Department of Physics, Durham University, Durham DH1 3LE, U.K.}
\author{Christian Arnold}
 \affiliation{
 Institute for Computational Cosmology, Department of Physics, Durham University, Durham DH1 3LE, U.K.}%
\author{Baojiu Li}
 \affiliation{
 Institute for Computational Cosmology, Department of Physics, Durham University, Durham DH1 3LE, U.K.}%

\date{\today}

\begin{abstract}
Future 21 cm intensity mapping surveys such as {\sc ska} can provide precise information on the spatial distribution of the neutral hydrogen (HI) in the post-reionization epoch. This information will allow to test the standard $\Lambda$CDM paradigm and with that the nature of gravity. In this work, we employ the {\sc shybone} simulations, which model galaxy formation in $f(R)$ modified gravity using the IllustrisTNG model, to study the effects of modified gravity on HI abundance and power spectra. 
We find that the enhanced growth low-mass dark matter halos experience in $f(R)$ gravity at high redshifts alters the HI power spectrum and can be observable through 21 cm intensity mapping. 
Our results suggest that the HI power spectrum is suppressed by $\sim13\%$ on scales $k\lesssim2\,h\,\mathrm{Mpc}^{-1}$ at $z=2$ for F6, a $f(R)$ model which passes most observational constraints. 
We show that this suppression can be detectable by {\sc ska}1-{\sc mid} with 1000 hours of exposure time, making HI clustering a novel test of gravity at high redshift.
\end{abstract}

\maketitle


\section{\label{sec:level1}Introduction}

Our standard model of cosmology -- the $\Lambda$ Cold Dark Matter model ($\Lambda$CDM) -- has proved very successful in describing almost all currently-available observational data of the Universe. Its underlying theory of gravity, Einstein's general relativity (GR), has been tested to remarkably high precision on small scales \citep{will2004}. In recent years, in the wake of high-precision astronomical observations, tests of GR on cosmological scales have become possible and commonplace as well \citep{koyama_MG_review}, although until now these tests have primarily focused on comparatively high-mass objects and low redshifts (e.g., \citep{terukina2014, cluster_constraints_1, cluster_constraints_2, cluster_constraints_3, cluster_constraints_4,  low_density_constraints_1, low_density_constraints_2, low_density_constraints_3, low_density_constraints_4, low_density_constraints_5}).
Due to the screening mechanisms which many alternatives to GR employ to pass the stringent Solar System tests, these objects are less suited to distinguish GR from alternative, \textit{modified gravity} (MG) theories with screening mechanisms.

As a representative example, we consider a particular one of these MG models in this paper -- $f(R)$ gravity \citep{buchdahl1970}, though we expect our conclusions to hold at least qualitatively for general thin-shell screening \cite{thin_shell_screening} models. $f(R)$ gravity is a generalization of GR which alters cosmic structure formation through a factor-of-$4/3$ enhanced gravitational force.
We adopt the popular variant proposed in Ref.~\cite{Hu:2007nk}, which, with certain choices of the free parameters of the model, can produce a cosmic expansion history very close to that of the $\Lambda$CDM paradigm. The model employs the so-called {\it chameleon screening} mechanism \cite{chameleon_1,chameleon_2} to ensure that the modifications to standard gravity are suppressed and GR-like behavior is recovered in high-density regions like the Solar System. The model considered here has been widely studied using numerical simulations, e.g., \citep{schmidt2010, zhao2011, li2011, lombriser2013, puchwein2013, arnold2014, low_density_constraints_3, arnold2018}. In this work we consider two instances of it \cite{Hu:2007nk}: F6 and F5, with its model parameter $f_{R0}$ equal to $-10^{-5}$ and $-10^{-6}$, respectively (see Section~\ref{sec:MGmodel} for more details). 

Although the constraints from our local environment are very tight, as mentioned above, the previous constraints from cosmological scales are much weaker since the objects used in these tests are generally more massive and well screened. One way to overcome this limitation is to study low-mass objects which are less likely to be screened and hence experience larger deviations from GR. However, a major challenge to this approach is the difficulty to accurately detect, resolve and trace such small objects in observations, even at low redshifts.

In this paper, we propose a novel test of gravity at intermediate scales and high redshifts ($z\geq 2$), using the distribution of neutral hydrogen (HI) in our Universe, which is observable in 21 cm experiments (some current and future instruments of this kind include {\sc ska} \cite{Paper-SKA1-MID}, {\sc m}eer{\sc kat} \cite{Santos:2017qgq}, {\sc lofar} \cite{2013A&A...556A...2V}, {\sc chime} \cite{2014SPIE.9145E..22B} and {\sc bingo} \cite{2013MNRAS.434.1239B}). 21 cm intensity mapping can be used to trace the underlying distribution of matter \citep{Bharadwaj:2000av,Loeb:2008hg,Bull:2014rha,Santos:2015gra} and with that the low-mass halos in the Universe (as suggested in \cite{Villaescusa-Navarro:2018vsg}).
In order to determine how possible deviations from GR would affect the HI distribution, we employ the \textsc{shybone} simulations, a set of full-physics hydrodynamical simulations of $f(R)$ modified gravity. 
Comparing the $f(R)$ simulations to their $\Lambda$CDM counterpart allows us to quantify the size of the MG effects on HI observables (such as the overall neutral hydrogen abundance and the HI power spectrum), and to assess if these effects can be observable with future 21 cm intensity mapping experiments. HI clustering has been proposed as a probe for a number of non-standard cosmological models, e.g., massive neutrinos, warm DM, dark energy and modified gravity \cite{21cm-non-st-phys_1, 21cm-non-st-phys_2, 21cm-non-st-phys_3, MF-non-stand-phys_1, MF-non-stand-phys_2}, but this study reveals new features, thanks to the high resolution of our simulations.

This paper {is structured} as follows. In Section~\ref{sec:ModelsandSim} we briefly introduce the two instances of Hu-Sawicki (HS) $f(R)$ gravity \cite{Hu:2007nk} used in our investigation and the suite of hydrodynamical simulations employed to quantify the abundance and clustering of HI. In Section~\ref{sec:Results}, we show and discuss the main results of this paper, including the overall neutral hydrogen density (\ref{sec:OverallHI}), the HI abundance in halos (\ref{sec:haloHIMF}) and the  HI power spectra in both real and redshift space (\ref{sec:ClusteringHI}).  Additional tests to explain the physics behind our results are performed in Section~\ref{sec:Addtests} and observational forecasts for a future 21cm intensity mapping experiment are discussed in Section \ref{sec:SKAres}.  In Section~\ref{sec:subgrid_approximation}, we comment on the dependence of our results on the galaxy formation model employed in our simulations. Finally, we conclude our findings in Section~\ref{sec:SummConc}.

\section{Theoretical models and simulations}
\label{sec:ModelsandSim}
\subsection{$f(R)$ gravity}
\label{sec:MGmodel}

$f(R)$ gravity is a popular class of MG models that is obtained by adding a scalar function $f(R)$ to the the Ricci scalar $R$ in the standard Einstein-Hilbert action \cite{buchdahl1970} of general relativity. With an appropriate choice of the functional form and parameters of $f(R)$, the theory can mimic the late time expansion history of a $\Lambda$CDM universe without explicitly having a cosmological constant $\Lambda$ (the accelerated expansion in these theories is achieved via some form of quintessence/cosmological constant and is not due to the modification of gravity itself \cite{Brax:2008hh,Wang:2012kj,Ceron-Hurtado:2016jrp}).

The action for the $f(R)$ gravity can be written as
\begin{equation}
S=\int \mathrm{d}^4x\, \sqrt{-g} \left[ \frac{R+f(R)}{16\pi G} +\mathcal{L}_m \right],\label{action}
\end{equation}
where $G$ is the gravitational constant, $g$ is the determinant of the metric, $g_{\mu\nu}$, and $\mathcal{L}_m$ is the standard matter/radiation Lagrangian density. The simulations considered here employ the weak-field and quasi-static limit (see \cite{weak-field_approx} for more details on the validity of these approximations), so that the equations of motion obtained by varying the action in Eq.~(\ref{action}) can be simplified to a (modified) Poisson equation plus an equation for the scalar degree of freedom, $f_{R} \equiv \d f(R)/\d R$ (the so-called {\it scalar} field),
\begin{eqnarray}\label{poissonfReq}
 \nabla^2 \Phi &=& \frac{16\pi G}{3}\delta\rho - \frac{1}{6} \delta R,\\
 \label{scalaronfReq}\nabla^2 f_{R} &=& \frac{1}{3}\left(\delta R -8\pi G\delta\rho\right),
\end{eqnarray}
where $\delta\rho \equiv \rho - \bar{\rho}$ and $\delta R \equiv R - \bar{R}$ are the matter density perturbation and the Ricci scalar perturbation, respectively (and $\bar{\rho}$ and $\bar{R}$ are their background values).

The HS variant of the theory \cite{Hu:2007nk} uses
\begin{equation}
 f(R) = -m^2\frac{c_1\left(\frac{R}{m^2}\right)^n}{c_2\left(\frac{R}{m^2}\right)^n +1},\label{fr}
\end{equation}
where $m$ is a new mass scale of the model, $m^2 \equiv \Omega_m H_0^2$, $H_0$ is the Hubble constant, $\Omega_{m}$ is the total non-relativistic matter energy density at present time in units of the present-day critical energy density of the Universe, $\rho_{c0}\equiv 3\,H^2_0/8\pi G$, and $c_1$, $c_2$, $n$ are model parameters. We choose $n=1$ hereafter for simplicity. 
Furthermore, if we tune the parameters $c_{1}$ and $c_2$ such that 
\begin{equation}
\frac{c_1}{c_2} = 6 \frac{\Omega_\Lambda}{\Omega_m} \quad \mathrm{and} \quad \frac{c_2|R|}{m^2} \gg 1,\label{conditions}
\end{equation}
the model leads to a cosmic expansion history which is very close to that of a $\Lambda$CDM universe \cite{Hu:2007nk}.    $\Omega_{\Lambda}$ in the above equation represents the cosmological constant energy density in units of $\rho_{c0}$ for the $\Lambda$CDM universe; in the case of $f(R)$ gravity, $\Omega_\Lambda$ still enters in the theory as a parameter. If one further assumes that the Universe is spatially flat, then $\Omega_\Lambda$ is simply given by $\Omega_\Lambda=1-\Omega_m$. This is our default assumption in this work.

The scalar field $f_R$ in this model can be approximated as
\begin{align}
f_{R} \equiv \frac{\mathrm{d}f(R)}{\mathrm{d}R} \approx -\frac{c_1}{c_2^2}\left(\frac{m^2}{R}\right)^{2},\label{fR}
\end{align}
and its background evolution can be expressed in terms of the background Ricci scalar $\bar{R}$,
\begin{align}
\bar{f}_{R}(a)  = \bar{f}_{R0} \left[ \frac{\bar{R}_0}{\bar{R}(a)} \right]^2,\label{fRa}
\end{align}
where $\bar{R}_0$ is the value of the Ricci scalar today and
\begin{align}
\bar{R}(a) = 3m^2\left[ a^{-3} + 4\frac{\Omega_\Lambda}{\Omega_m} \right]. \label{R_bar}
\end{align}
The theory is therefore fully specified by $\Omega_m$ and the present-day value of the background scalar field, $\bar{f}_{R0}$. 

In order to satisfy the stringent constraints on the possible deviations from standard gravity in our local environment \cite{will2004}, the theory employs the above-mentioned chameleon screening mechanism \cite{chameleon_1, chameleon_2} to suppress modifications to gravity and restore GR in high-density regions. The chameleon screening has been described in great detail in the literature and thus we will not discuss it further here, but instead simply mention that it becomes effective when $f_R$ becomes close to zero, such that $\delta R\approx8\pi G\delta\rho$ according to Eq.~(\ref{scalaronfReq}), and then Eq.~(\ref{poissonfReq}) reduces to the standard Poisson equation in Newtonian gravity. The screening is more likely to take place at earlier times when matter density is high and the background value of the scalar field, $|\bar{f}_R|$, is small. At a given time, this mechanism screens regions where their density is high and therefore the Newtonian potential is deep. 
The transition between screened and unscreened regimes depends on the choice of $\bar{f}_{R0}$. In $f(R)$ gravity, the speed of gravitational wave (GW) is equal to the speed of light and the model passes recent constraints from GW observations \cite{GW}, making it one of the most promising alternatives to GR. 

As mentioned in the introduction, in this work we will focus on the F5 and F6 instances of HS $f(R)$ gravity, for which $\bar{f}_{R0}$ is equal to $-10^{-5}$ and $-10^{-6}$, respectively. In F5, the effects of MG are stronger than in F6 and this model is now ruled out by observational constraints \cite{terukina2014} (see \cite{Burrage:2017qrf} for a review about the recent constraints on chameleon gravity). However, it is used here as a toy model to assess the effects of a stronger deviation from GR on the HI distribution. The F6 model is consistent with most cosmological observations. 

An important characteristic of the chameleon screening is that this mechanism becomes inefficient for small structures at high redshift, while more massive objects and denser environments  become unscreened at later times (lower redshift). At high redshift, low-mass halos are already unscreened and affected by modified gravity. As a consequence, by observing such small structures we can, in principle, place constrains on the $f(R)$ deviations from GR. As we will see in the next sections, 21 cm intensity mapping is sensitive to the abundance of halos down to $10^9\,\mathrm{M}_\odot$,  making it a very promising probe of differences at the low-mass end of the halo mass function, without the need to resolve individual halos.

\subsection{Full-physics simulations in MG}
\label{sec:Shybonesim}

In order to quantify how modifications to gravity affect the 21 cm signal, we analyze the {\sc shybone} simulations \citep{arnold2019}, a set of high-resolution full-physics hydrodynamical simulations of HS $f(R)$ gravity, carried out with the moving mesh simulation code \textsc{arepo} \citep{2010MNRAS.401..791S}. The suite includes two subsets of simulations: a large-box set with a box size of $L=62\, h^{-1} \text{Mpc}$ ({S}62 hereafter) and a small-box set for which $L=25\, h^{-1} \text{Mpc}$ ({S}25 hereafter), both with roughly $2\times 512^3$ resolution elements ($h$ is the dimensionless Hubble constant, given by $h\equiv H_0/(100\,{\rm km}\,{\rm
s}^{-1}{\rm Mpc}^{-1})$). The {S}62 simulations have a mass resolution of  $m_{\rm DM} = 1.3 \times 10^8\, h^{-1} \,\mathrm{M}_\odot$ for DM-particles and roughly $m_{\rm gas} = 2.5 \times 10^7\, h^{-1}\, \mathrm{M}_\odot$ for gas cells, and they have been run for GR, F6 and F5 up to $z=0$. The {S}25 simulations have a mass resolution of $m_{\rm DM} = 8.4 \times 10^6\,h^{-1}\,\mathrm{M}_\odot$ and $m_{\rm gas} = 1.6 \times 10^6\, h^{-1}\,\mathrm{M}_\odot$ and have been run for the same three models, up to $z=0$ for GR and F6 and up to $z=1$ for F5 (the enhanced gravitational interactions in the F5 model considerably increase the computational cost of the simulations compared to their GR counterpart). {S}25 features a higher resolution, but its smaller box means that we inevitably lose some information of large-scale modes and massive halos (see the discussion in the next section). The {S}62 suite features also
DM-only (DMO hereafter) counterparts for all the runs, which are used to compare the halo mass function from full-physics and DMO simulations below. All simulations adopt the Planck 2016 \citep{planck2016} cosmology with $\Omega_m = 0.3089$, $\Omega_B = 0.0486$, $\Omega_\Lambda = 0.6911$, $ h = 0.6774$, $\sigma_8 = 0.8159$ and $n_{\rm s} = 0.9667$, where $\Omega_B$ is the present-day baryon density parameter, $\sigma_8$ is the root-mean-squared matter density fluctuation over spherical regions with radius $8h^{-1}$Mpc at $z=0$, and $n_s$ is the index of the primordial power spectrum.

The full-physics simulations use the {IllustrisTNG} hydrodynamical model \citep{pillepich2018,pillepich2018b,springel2018,Vogelsberger2014,vogelsberger2014b,weinberger2017,genel2018,marinacci2018, nelson2018}, incorporating a prescription of star and black hole formation and feedback, gas cooling, galactic winds and magneto-hydrodynamics on a moving Voronoi mesh \citep{pillepich2018, weinberger2017}. The equations for $f(R)$ gravity are solved to full non-linearity in the Newtonian limit by the modified gravity solver in the code \citep{arnold2019}, fully capturing the effects of the chameleon screening.

To calculate the neutral hydrogen fraction in each Voronoi cell, we follow  the prescription in Section~2.2 in  \cite{Villaescusa-Navarro:2018vsg}. For non star-forming gas, we use the neutral hydrogen fraction calculated on-the-fly in the simulations, while for star-forming gas we post-process the outputs, recalculating the neutral hydrogen fraction in each cell assuming a temperature of $T=10^4\,\mathrm{K}$ and following the approach in \cite{2013MNRAS.430.2427R} to take into account self-shielding corrections. The post-processing gives the total fraction of hydrogen that is non-ionized: atomic (HI) and molecular hydrogen (H$_2$). Because we are solely interested in HI, we calculate and subtract the fraction of $\mathrm{H}_2$ for each Voronoi cell as in \cite{Villaescusa-Navarro:2018vsg}. 

\section{Results}
\label{sec:Results}
\subsection{Overall neutral hydrogen density}
\label{sec:OverallHI}
\begin{figure}
\advance\leftskip-0.6cm
  \includegraphics[width=.59\textwidth]{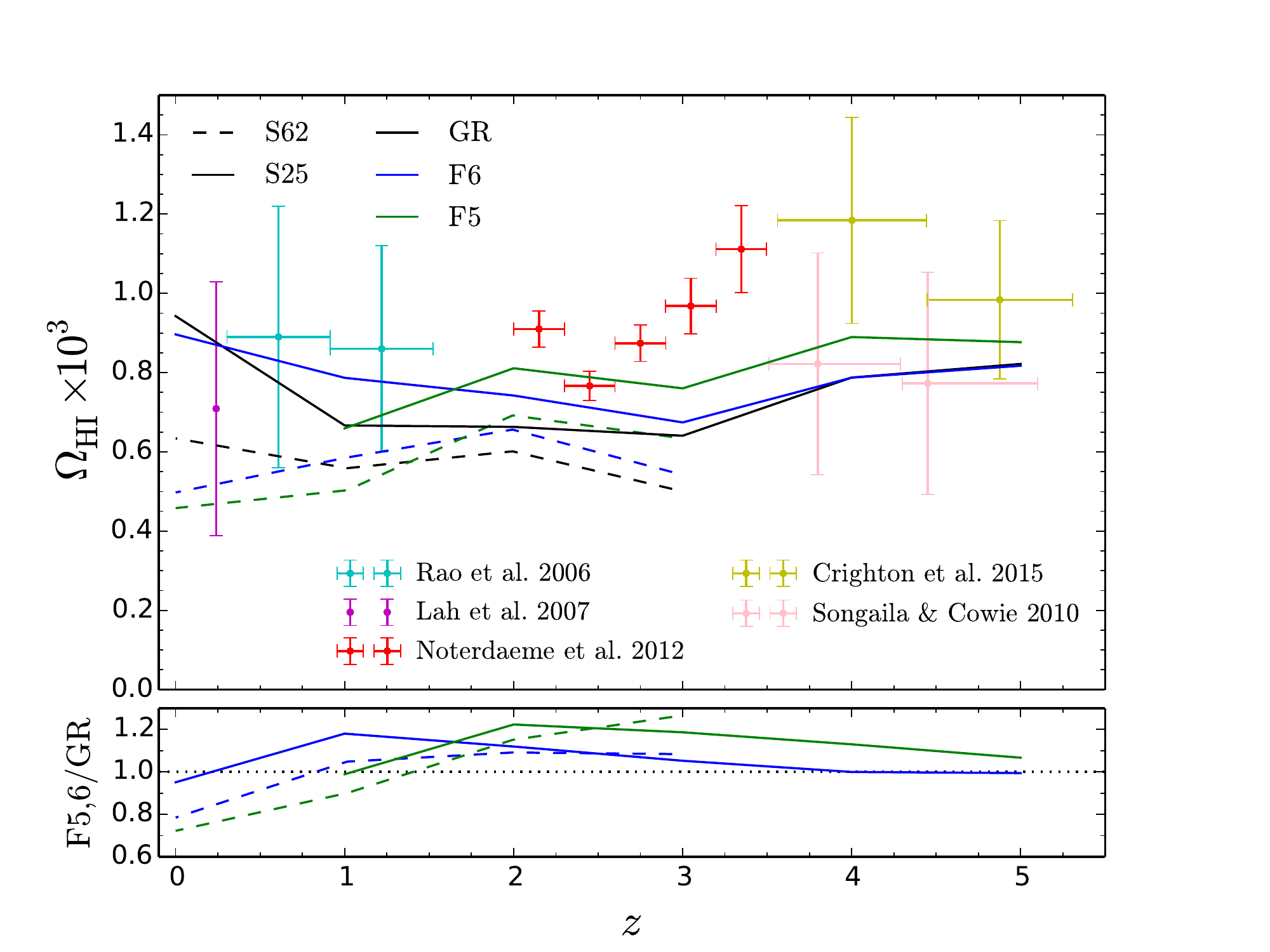}
  \caption{{\it Top panel}: Overall HI abundance, $\Omega_\mathrm{HI}(z) = \bar{\rho}_\mathrm{HI}(z)/\rho_\mathrm{c0}$, where $\bar{\rho}_\mathrm{HI}(z)$ is the mean HI density, from GR (black), F6 (blue) and F5 (green), compared with observationally measured values (symbols). Solid lines refer to {S}25 simulations, while dashed lines to {S}62 simulations. {\it Bottom panel}: the relative differences of the simulation predictions from F6 (blue) and F5 (green) w.r.t. GR.}
\label{fig:Fig1} 
\end{figure}

In Fig.~\ref{fig:Fig1} we show the overall neutral hydrogen density measured from the {S}62 (in the range $0\leq z\leq3$ for all models) and {S}25 (in the range $0\leq z\leq5$ for GR and F6 and $1\leq z\leq5$ for F5) simulations. We follow the common definition for the overall HI abundance, $\Omega_\mathrm{HI}(z) = \bar{\rho}_\mathrm{HI}(z)/\rho_\mathrm{c0}$, where $\bar{\rho}_\mathrm{HI}(z)$ is the mean HI density in our simulations at a given redshift $z$ and $\rho_\mathrm{c0}$ is the present-day critical density as defined above. We also show a selection of observational data for the HI abundance at different redshifts from \cite{Rao:2005ab,Lah:2007nk,2010ApJ...721.1448S,2012A&A...547L...1N,Crighton:2015pza}. 

First, we note that in  Fig.~\ref{fig:Fig1}  the HI abundance (for each model) predicted by {S}25 is higher than that measured from the low-resolution {S}62. A similar effect was found in \cite{Villaescusa-Navarro:2018vsg} comparing the low- and high-resolution TNG simulations. This discrepancy between simulations at different resolution can be understood as follows. The neutral hydrogen in the post-reionization epoch is concentrated in halos, where shielding effects screen them from ionization. It was shown (see, e.g., \cite{Villaescusa-Navarro:2018vsg}; but see also next section) that there is a significant amount of HI in halos with masses as low as $10^9\,\mathrm{M}_\odot$ at $z\leq 5$. This implies that resolving halos of this mass in simulations is essential to measure the HI abundance accurately. However, because of its lower resolution, {S}62 does not fully resolve halos with masses $<10^{10}\,\mathrm{M}_\odot$, and therefore predicts a lower value for the HI abundance at all redshifts considered in this analysis. On the other hand, {S}25 can resolve halos  down to $6\times 10^8\,\mathrm{M}_\odot$, and therefore produces more reliable results of the HI abundance. However, we note that because of their small box size, we do not have a statistically robust sample of halos with masses $\gtrsim10^{12}\,\mathrm{M}_\odot$ in {S}25. This explains why our high-resolution simulations predict slightly lower HI abundance at $z\leq 5$ compared to that measured in \cite{Villaescusa-Navarro:2018vsg} for GR using the TNG-100 simulation (performed in a box of co-moving length $L=75\,h^{-1}\,\mathrm{Mpc}$ at the same resolution as {S}25).

Considering the differences between $f(R)$ gravity and GR, in {S}25 the ratios of $\Omega_{\rm HI}(z)$ w.r.t. GR (solid lines in the lower panel of Fig.~\ref{fig:Fig1}) show a similar trend in F6 and F5. Indeed, in both models, $\Omega_\mathrm{HI}$ is similar to GR at high redshifts,  larger than GR at intermediate redshifts (with an enhanced peak at $z=1$ and $z=2$ for F6 and F5, respectively) and falls below the GR values for $z<1$. Overall, HI is $\sim 5\%$ ($18\%$) more abundant in F6 (F5) than in GR at $z=3$. At $z=2$ there is $\sim 12\%$ ($22\%$) more HI in F6 (F5) than GR, while at $z=1$ we find more HI in F6 than in F5 and GR. This behavior can be understood as follows. At high redshifts ($z\gtrsim4-5$), {modified gravity effects} on the matter and halo distribution are screened for the models considered here, and thus F5 and F6 both behave similarly to GR. At intermediate redshifts ($z=2\sim3$), {low-mass} halos in F5 and F6 become unscreened and experience enhanced growth, leading to increased abundance of these low-mass objects compared to GR. Since neutral hydrogen can survive only in self-shielding halos, this implies that in F5 and F6 there are more HI-hosting halos than in GR, and, consequently, these models are characterized by larger overall HI abundance. At low redshifts, baryonic effects become important. At these redshifts, we suspect that processes of gas heating can be more efficient in MG than in GR, reducing the overall HI abundance in F5 and F6 compared to GR. A closer inspection of Fig.~\ref{fig:Fig1} suggests that F6 behaves as a `retarded' version of F5, with the maximum enhancement w.r.t.~GR shifted to lower redshifts. This is expected as the screening is more efficient in F6 and thus leads to a later onset of the MG force enhancement compared to F5.

\subsection{HI mass in halos}
\label{sec:haloHIMF}

\begin{figure*}
    \centering
\subfigure[\quad $z=2$, GR]{    \includegraphics[width=.33\textwidth]{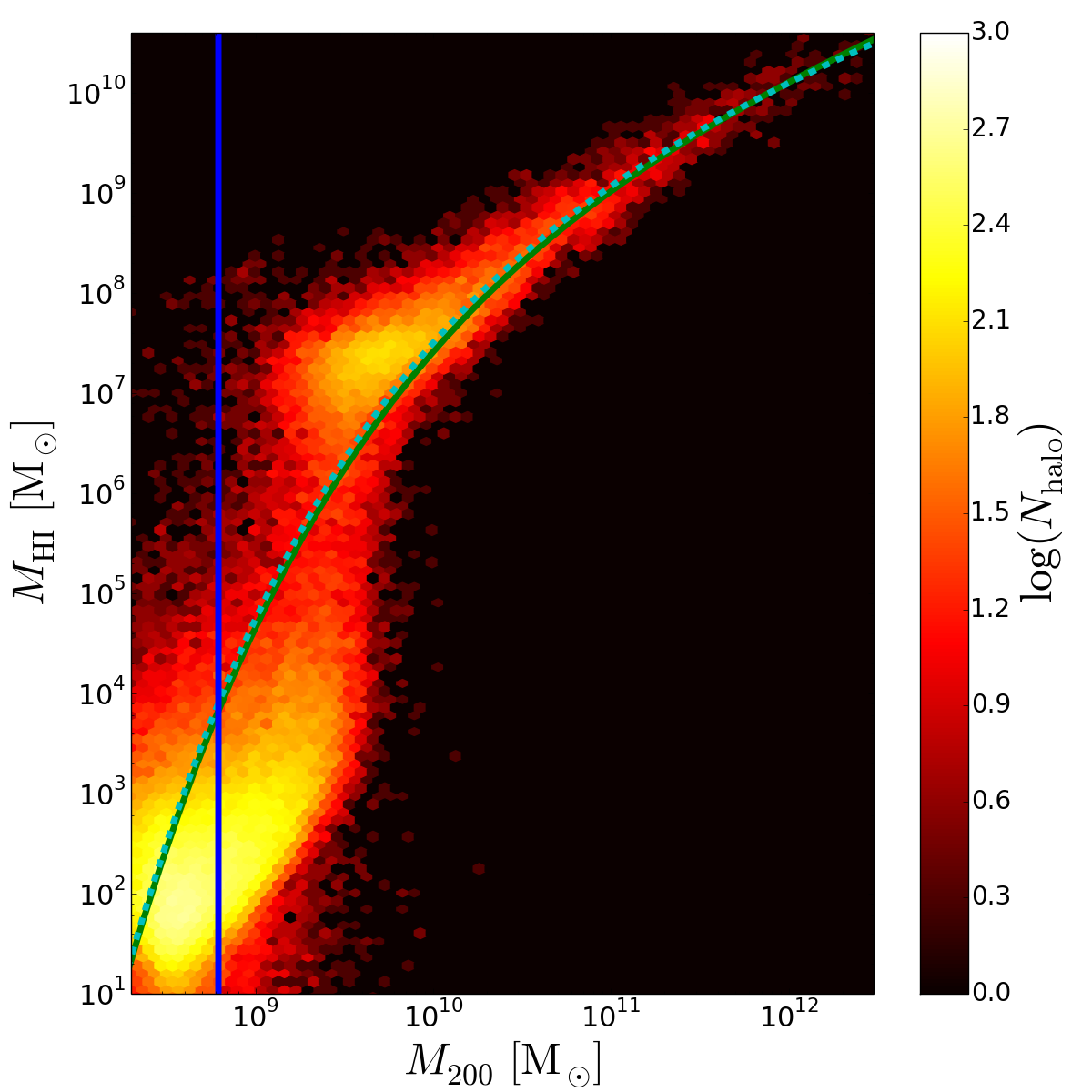}
}\hspace{-2.ex}
\subfigure[\quad $z=2$, F6]{    \includegraphics[width=.33\textwidth]{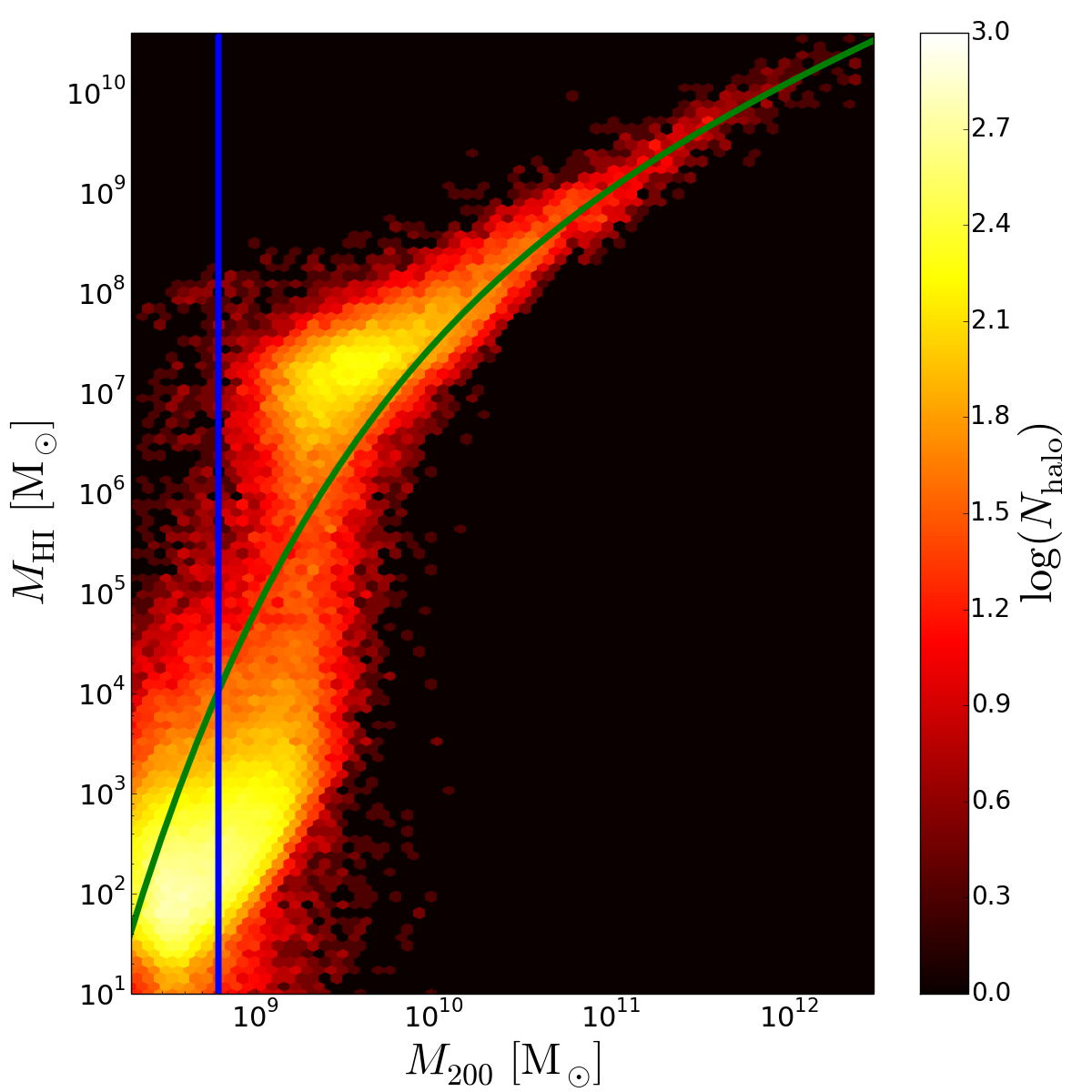}
}\hspace{-2.ex}
\subfigure[\quad $z=2$, F5]{    \includegraphics[width=.33\textwidth]{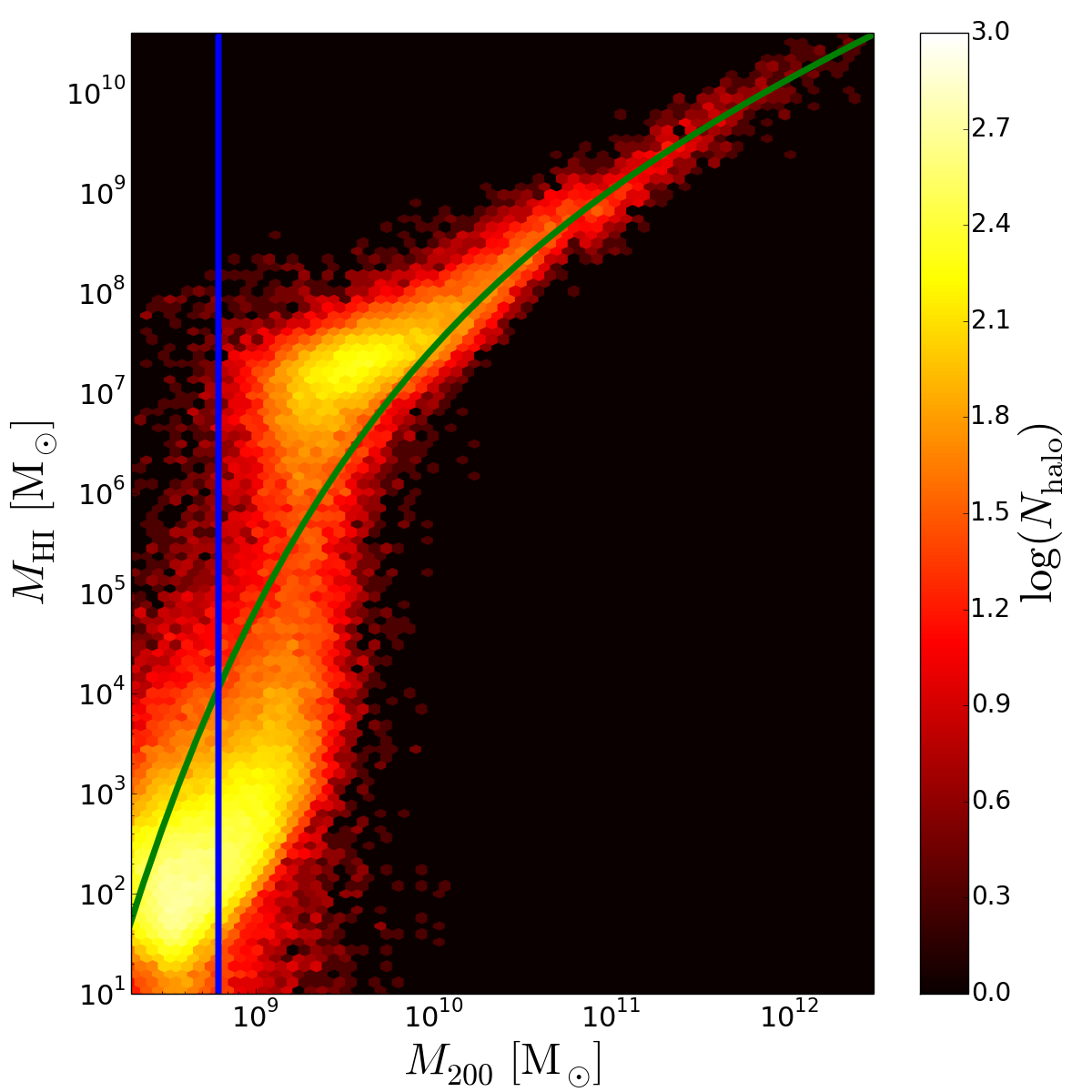}
}
\subfigure[\quad $z=3$, GR]{    \includegraphics[width=.33\textwidth]{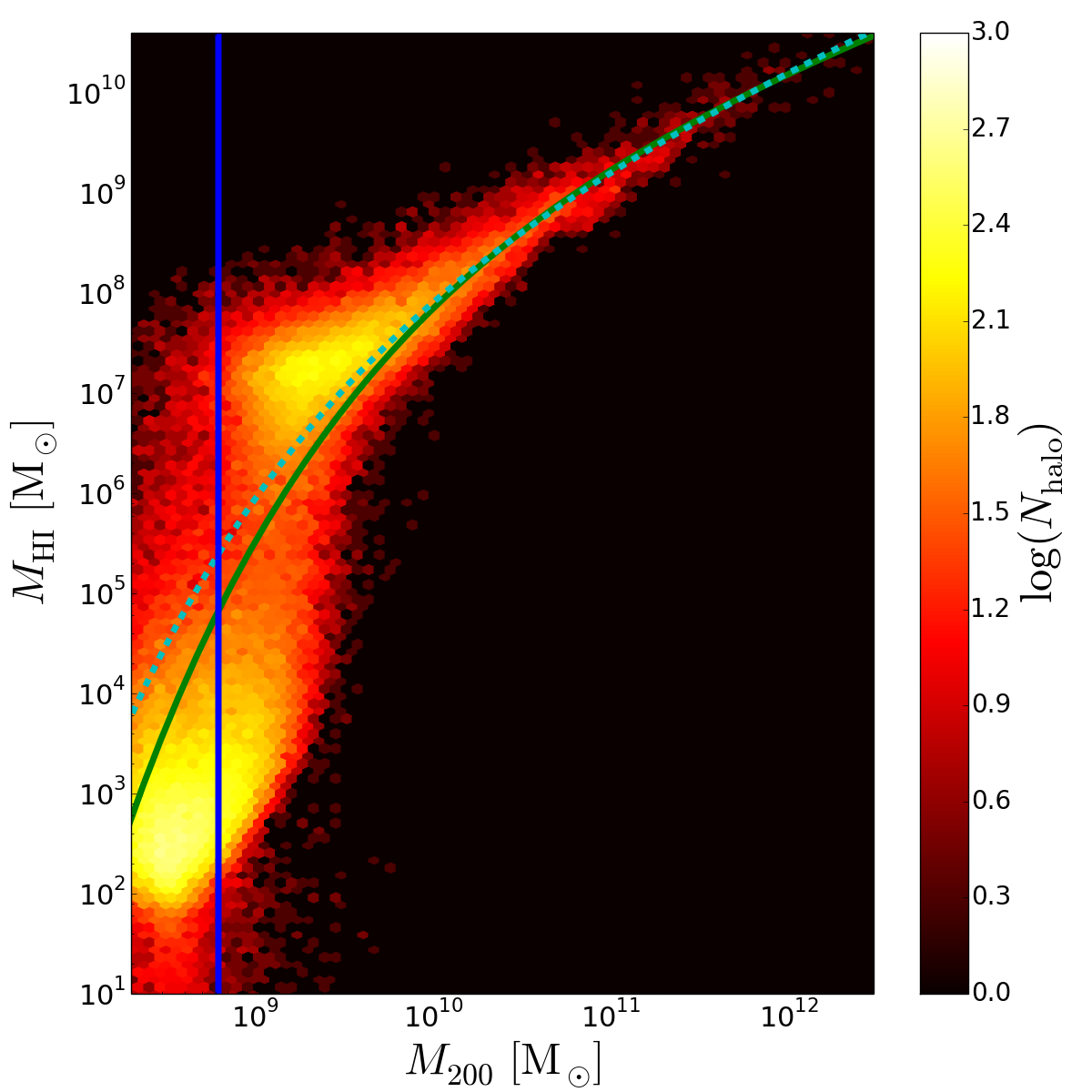}
}\hspace{-2.ex}
\subfigure[\quad $z=3$, F6]{    \includegraphics[width=.33\textwidth]{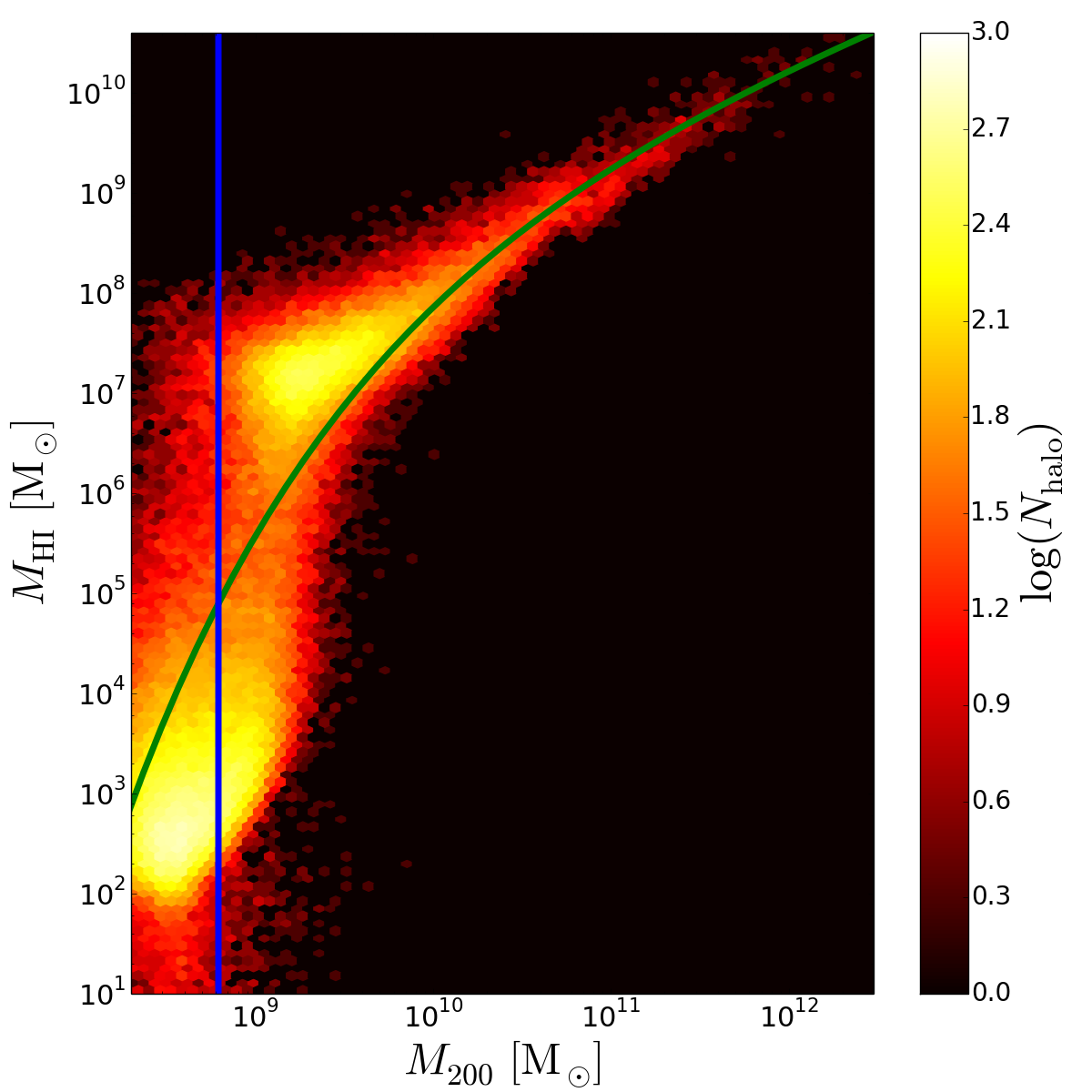}
}\hspace{-2.ex}
\subfigure[\quad $z=3$, F5]{    \includegraphics[width=.33\textwidth]{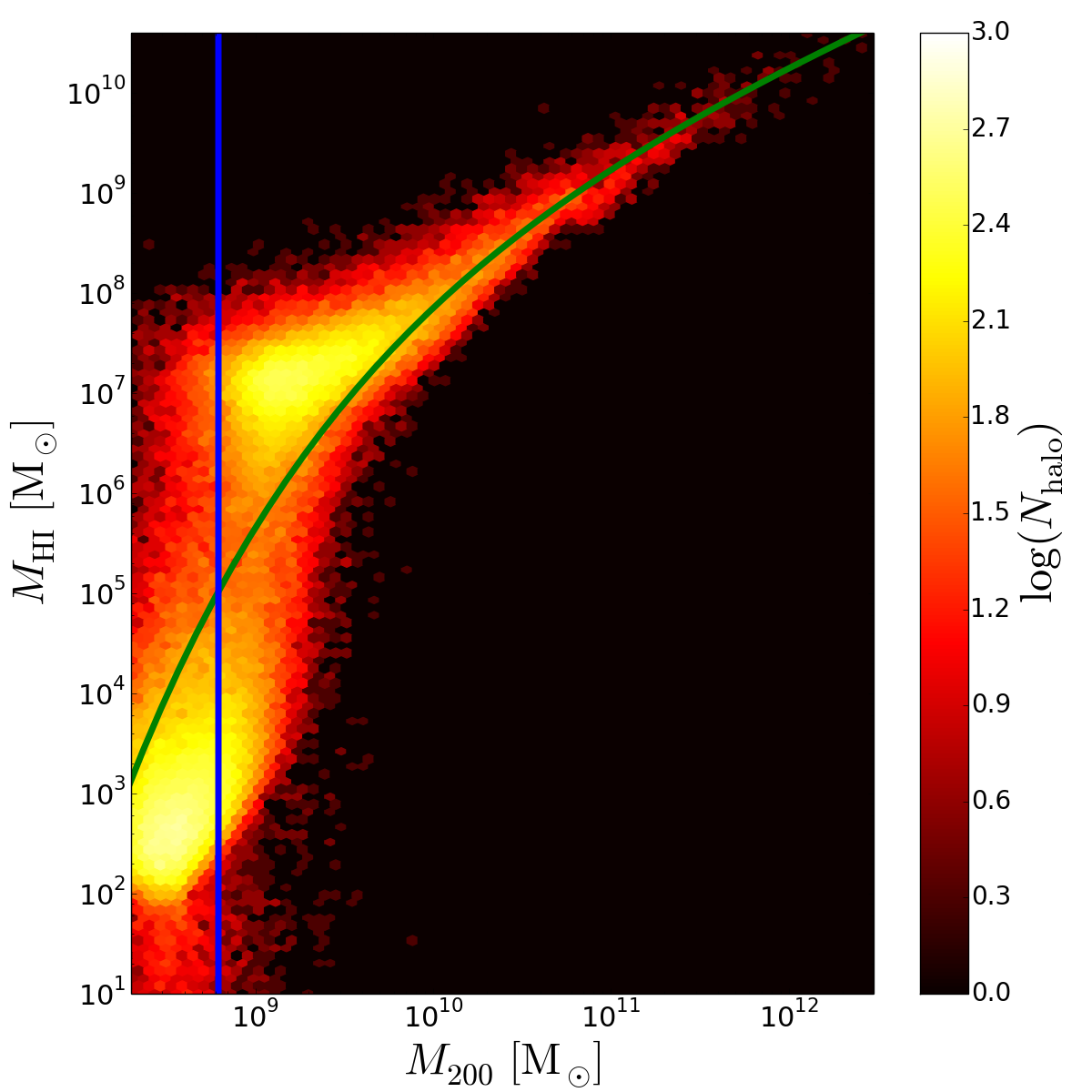}
}
\caption{HI mass contained in each halo at $z=3$ and $z=2$ measured from the {S}25 simulations for GR, F6 and F5 respectively (as labelled). The green solid line in each panel shows the best-fit curve obtained from Eq.~(\ref{eq:fitting}). The best-fit values of the free parameters are displayed in Table~\ref{table:fit}. The cyan dotted line indicates the best-fit curve of Eq.~(\ref{eq:fitting}) found in \cite{Villaescusa-Navarro:2018vsg} for GR.  The blue vertical line indicates the mass of halos with around 50 DM (simulation) particles in {S}25.}
    \label{fig:haloHImassf}
\end{figure*}

In this subsection, we present and discuss the {\it halo HI mass function}, i.e., the average HI mass enclosed in halos as a function of the halo mass. 
In the post-reionization epoch, the majority of HI resides in halos, and therefore an accurate knowledge of the halo HI mass function can be used to predict the HI power spectrum in real and redshift space without requiring the full hydrodynamical simulation apparatus, but instead by painting the HI on top of dark matter halos from DMO simulations \cite{Villaescusa-Navarro:2014cma}.  This approximation has been applied to extract information on the HI power spectrum for non-standard cosmological models of dark matter and dark energy \cite{21cm-non-st-phys_1, 21cm-non-st-phys_2, 21cm-non-st-phys_3}.   

Here, for each halo in our simulations we measure its enclosed HI mass. The halos 
are identified using the \textsc{subfind} \citep{Springel:2000qu} algorithm implemented in \textsc{arepo}. The halo mass $M_\mathrm{halo}$ is defined as $M_{200}$, the mass contained in a sphere of radius $r_{200}$, within which the average density is $200$ times the critical density at the specified redshift. In Fig.~\ref{fig:haloHImassf}, we plot the HI mass  as a function of the host halo mass for halos in the {S}25 simulations. As the figure shows, the HI mass increases monotonically with the halo mass on average in all models considered in our analysis, which is in agreement with what was found in \cite{Villaescusa-Navarro:2018vsg}. The plot also shows that HI is present within very low-mass halos, $\sim 10^9\,\mathrm{M}_\odot$, which highlights  the importance of the simulation resolution for HI clustering estimates.   

In \cite{Villaescusa-Navarro:2018vsg}, the authors identified a fitting function,
\begin{equation}
M_\mathrm{HI}(M_{200}, z) = M_0 \,\left(\frac{M_{200}}{M_\mathrm{min}} \right)^\alpha\, \exp\left[-\left(\frac{M_\mathrm{min}}{M_{200}}\right)^{0.35}\right],
\label{eq:fitting}    
\end{equation}
which captures the power-law behavior at high halo masses, $M_\mathrm{HI}\propto M_{200}^\alpha$ and the exponential cut-off for masses $M_{200}\lesssim M_\mathrm{min}$. We fit our data with this formula by dividing the halo mass range $\log(M_{200}/\mathrm{M}_\odot)\in [8.0,12.5]$ into two sub-ranges: $[8.0,11.0]$ and $[11.0,12.5]$. In order to account for the fact that at larger masses the halo sample is much smaller, we divide these two sub-ranges into 40 and 9 bins respectively, calculate the mean and variance of $\log(M_\mathrm{HI}/\mathrm{M}_\odot)$ in each bin, and use a minimum-$\chi^2$ method to obtain the best-fit parameters. We find that this function remains a good fit to our simulations for all the models studied here (see Fig.~\ref{fig:haloHImassf}). The best-fit values for the free parameters $\{M_0,\alpha,M_\mathrm{min}\}$ are shown in Table~\ref{table:fit} for $z=2$ and $z=3$ and for GR, F6 and F5.

\begin{table}
\begin{tabular}{c|c|c|c|c} 
  \hline\hline
\quad\quad \quad & \quad$z$\quad\quad &  \quad$M_0$\quad\quad&  \quad$\alpha$\quad\quad&  \quad$M_\mathrm{min}$\quad\quad \\ 
  \quad Model\quad\quad & \quad\quad\quad &  \quad$ [\mathrm{M}_\odot]$\quad\quad&  \quad\quad\quad&  \quad$ [\mathrm{M}_\odot]$\quad\quad \\ 

  \hline
  GR &  $3.0$ & $1.30\times 10^{10}$ & $0.62$ & $2.72\times 10^{11}$ \\
  

 \hline
  F6 & $3.0$ & $1.19\times 10^{10}$ & $0.65$ &  $2.40\times 10^{11}$ \\ 
  \hline
 F5 & $3.0$ & $8.81\times10^{9}$  & $0.72$ & $1.79\times 10^{11}$   \\
   \hline
  GR & $2.0$ & $1.69\times 10^{10}$ & $0.67$ & $4.80 \times 10^{11}$  \\ 

 \hline
  F6 & $2.0$ & $1.52\times10^{10}$ & 0.66& $4.25\times 10^{11}$\\ 
  \hline
 F5 & $2.0$ & $1.34\times 10^{10}$ & $0.73$ & $3.65\times 10^{11}$  \\
  \hline\hline
\end{tabular}
\caption{Best-fit values of the free parameters for the Eq.~(\ref{eq:fitting}) at $z=2$ and $3$ for GR, F6 and F5 (see main text for details).}
\label{table:fit}
\end{table}

In Fig.~\ref{fig:haloHImassf},
we also compare our best-fit curves with the GR results taken from 
\cite{Villaescusa-Navarro:2018vsg} (cyan dotted lines). As we can see, the fitting results in the two works agree very well at $z=2$ for the entire halo mass range. At $z=3$, our result is again in good agreement with \cite{Villaescusa-Navarro:2018vsg} for $M_{200}\gtrsim5\times 10^{9}\,\mathrm{M}_\odot$, while
slightly disagreeing at lower masses. This can be due to differences in the details of the fitting procedure or due to cosmic variance because of our smaller simulation box.

A quick comparison of our results with Fig.~4 in \cite{Villaescusa-Navarro:2018vsg} shows that, again because of the smaller box size, our high-resolution simulations do not contain halos more massive than $\sim 10^{12}\,\mathrm{M}_\odot$. This can explain why our total HI abundance (see \ref{sec:OverallHI}) is slightly lower than that found in \cite{Villaescusa-Navarro:2018vsg}.
It is nevertheless possible to correct the total HI abundance in our simulations for the missing contribution from high mass halos, $\Omega^{\rm corr}_\mathrm{HI}$. Given the expression for the total $\Omega^{\rm tot}_\mathrm{HI}$ in the Universe (and assuming that all HI is contained in halos):
\begin{equation}
\Omega^{\rm tot}_\mathrm{HI} =\frac{1}{\rho_\mathrm{c0}} \int^\infty_{0} M_\mathrm{HI}(M_{200})\frac{{\rm d}n_\mathrm{halo}}{{\rm d}M_{200}}{\rm d}M_{200},
\end{equation}
in which ${\rm d}n_\mathrm{halo}/{\rm d}M_{200}$ is the halo mass function (HMF) and $M_\mathrm{HI}(M_{200})$ is given by Eq.~(\ref{eq:fitting}), the missing HI in the simulations can be estimated as 
\begin{equation}
\Omega^{\rm corr}_\mathrm{HI} =\frac{1}{\rho_\mathrm{c0}} \int^\infty_{M^{\rm cut}_{200}}M_\mathrm{HI}(M_{200})\frac{{\rm d}n_\mathrm{halo}}{{\rm d}M_{200}}{\rm d}M_{200}.
\label{eq:Omega_HI_corr}
\end{equation}
Here $M^{\rm cut}_{200}$ is the maximum halo mass in the considered simulation, while the HMF is estimated using the approach of \cite{Sheth:1999su}.  Eq.~(\ref{eq:Omega_HI_corr}) gives an approximate estimate for the HI fraction in large halos which are missing in {S}25, $\Omega^{\rm corr}_\mathrm{HI} \approx 7\times10^{-5}$ at $z=3$ for GR. Adding this contribution to the $\Omega_\mathrm{HI}$ measured from our simulations (cf.~Fig.~\ref{fig:Fig1}) brings the total $\Omega_\mathrm{HI}$ in very good agreement with that found in \cite{Villaescusa-Navarro:2018vsg}.

As mentioned already, the above fitting  results for the halo HI mass function can be used to model the HI distribution in DMO simulations, without having to run computationally-expensive high-resolution hydrodynamical simulations, but using a {\it halo occupation distribution} (HOD) technique applied to HI (see e.g., \cite{Villaescusa-Navarro:2014cma} for more details).

\subsection{HI clustering}
\label{sec:ClusteringHI}

In the previous subsections, we have focused on the overall HI abundance and the halo HI mass function. Although these quantities {provide} useful information about the total HI in our simulation boxes and the HI inside halos, they do not directly describe the HI distribution and clustering. To understand the differences in the matter clustering of HI, we now analyze the HI power spectrum, $P_{\rm HI}(k)$. 

\begin{figure*}
\advance\leftskip-0.5cm
  \includegraphics[width=1.05\textwidth]{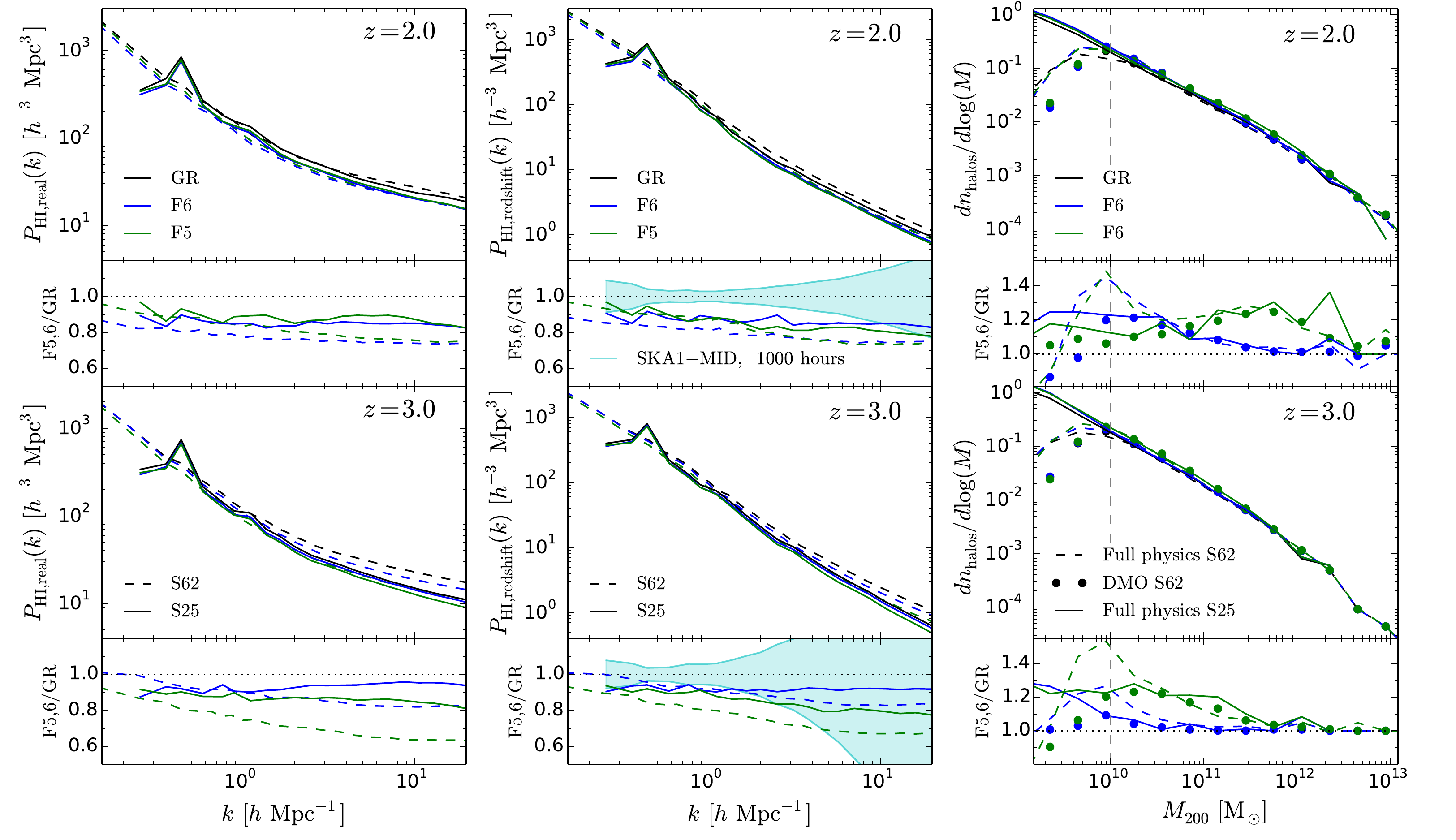}
  \caption{Neutral hydrogen power spectra and halo mass functions at $z=2$ (upper panels) and $z=3$ (lower panels) for GR (black), F6 (blue) and F5 (green). {\it Left panels}: Real space HI power spectra $P_{\rm HI}$. Dashed lines show the results for {S}62, while solid lines show those for {S}25. The real-space power spectra are taken from Ref.~\cite{arnold2019}. {\it Central panels}: Same as left panels but for the redshift space (monopole) HI power spectra. The cyan shaded areas in the lower sub-panels  represent the expected errors for {\sc ska}1-{\sc mid} measurements for GR, 
  assuming $1000$ observing hours. The error bars are calculated using the GR redshift space power spectrum measured from {S}25. {\it Right panels}: Differential halo mass functions as shown in \cite{arnold2019b}, for comparison. Solid (dashed) lines show the results from full-physics {S}25 ({S}62)  simulations. Symbols indicate the results from DMO {S}62 simulations. The lower subpanels show the relative differences to GR.}
\label{fig:Fig_HIPK_HMF}
\end{figure*}

We calculate $P_\mathrm{HI}(k)$ in redshift space from the simulation outputs by applying a cloud-in-cell (CIC) density assignment scheme (\cite{He:2015bua,Jing:2004fq}), while the HI power spectra in real space from \cite{arnold2019} (these results were obtained from the same simulation suite) are shown for comparison. In the left and central panels of Fig.~\ref{fig:Fig_HIPK_HMF}, we display the measured power spectra for the HI over-density, $\rho_\mathrm{HI}/\bar{\rho}_\mathrm{HI}$, where $\bar{\rho}_\mathrm{HI}$ is the mean HI density defined above. We present the $P_\mathrm{HI}(k)$ results measured from both the {S}62 (dashed lines) and {S}25 (solid lines) simulations. The overall effect in F6 and F5 (seen in both the low- and high-resolution simulations and in both real and redshift space) is a scale-dependent reduction of the clustering power w.r.t. GR. However, the results for the high- and low-resolution simulations do not converge at the scales probed by the simulations. As mentioned above, the disagreement in the results from {S}62 and {S}25 reflects the lack of resolved small halos with $M_{200}<10^{10}\,\mathrm{M}_\odot$ in the former. For this reason, we will only discuss the results from the {S}25 simulations in the following. 

For the monopole of the redshift space power spectrum, we find that at $z=3$ the F6 (F5) $P_\mathrm{HI,\, redshift}(k)$ is suppressed by $8\%$ ($14\%$) for $k\sim 2\,h\,\mathrm{Mpc}^{-1}$, while the effect is even stronger at higher wave numbers. At $z=2$, for F6 (F5) it is suppressed by $13\%$ ($18\%$) w.r.t. GR for $k\sim 2\,h\,\mathrm{Mpc}^{-1}$.  
Similar trends can be found for the real-space power spectrum. It is interesting to note that at $z=2$ the F6 $P_\mathrm{HI,\,real}(k)$ is slightly more suppressed than that for F5 at all scales probed by our simulations, while at $z=3$ the suppression is stronger for F5. 

To understand the above results, in the right panels of Fig.~\ref{fig:Fig_HIPK_HMF}, we compare to the HMFs of the {\sc shybone} simulations (taken from \citep{arnold2019b}). In the case of {S}62, we show the HMFs for both the full-physics (dashed lines) and DMO (symbols) simulations, while for {S}25 we only show the full-physics ones (solid lines). The ratios relative to GR for F5 and F6 measured from full-physics {S}62 and full physics {S}25 agree  very well for halos with masses $>3\times 10^{10}\,\mathrm{M}_\odot$. However, due to the lower resolution of {S}62 (halos with $M_{200}\sim 10^{10}\,\mathrm{M}_\odot$ contain roughly 50 DM particles in this box), the HMFs disagree at lower masses. Since halos with $10^9\,\mathrm{M}_\odot < M_{200}< 10^{10}\,\mathrm{M}_\odot$ can host appreciable amounts of HI \cite{Villaescusa-Navarro:2018vsg},  this explains why the ratios of the $f(R)$ HI power spectra w.r.t. GR for the large box do not agree with those measured from the small box.  

Analyzing the behavior of the HMF ratios, we find that at $z\geq2$ F5 and F6 are characterized by a larger number of low-mass halos ($M_{200}\lesssim10^{12}\,{\rm M}_\odot$) than GR. As HI can survive only in halos where self-shielding effects prevent it from ionization, in MG there are more hosts for HI than GR. Therefore, our interpretation of the behavior of the HI power spectra in the different models is that it primarily reflects the differences in the HMFs of these models (though these models also have different halo density profiles \cite{Arnold:2014qha}, which can have impact on the HI distribution as well), with $f(R)$ gravity being able to turn more low initial density peaks into halos. Given that low initial density peaks are less clustered, HI, as a tracer of them, has a smaller clustering amplitude in $f(R)$ models compared to GR.

Comparing the HMFs measured from the DMO and the full physics simulations in the right column of Figure \ref{fig:Fig_HIPK_HMF}, it is obvious that the galaxy formation processes have a non-negligible effect on the HMF itself. The relative differences in the HMFs between the $f(R)$ models and GR (lower panels in the right column) are nevertheless only mildly affected by the baryons for $M_{200}\gtrsim 3\times 10^{10}\,\mathrm{M}_\odot$ at $z=2$ and $3$. At lower masses, we suspect that the differences between the full physics and DMO results are due to resolution effects.

\subsection{Additional tests}
\label{sec:Addtests}

\begin{figure}
\advance\leftskip-0.2cm
\includegraphics[width=0.47\textwidth]{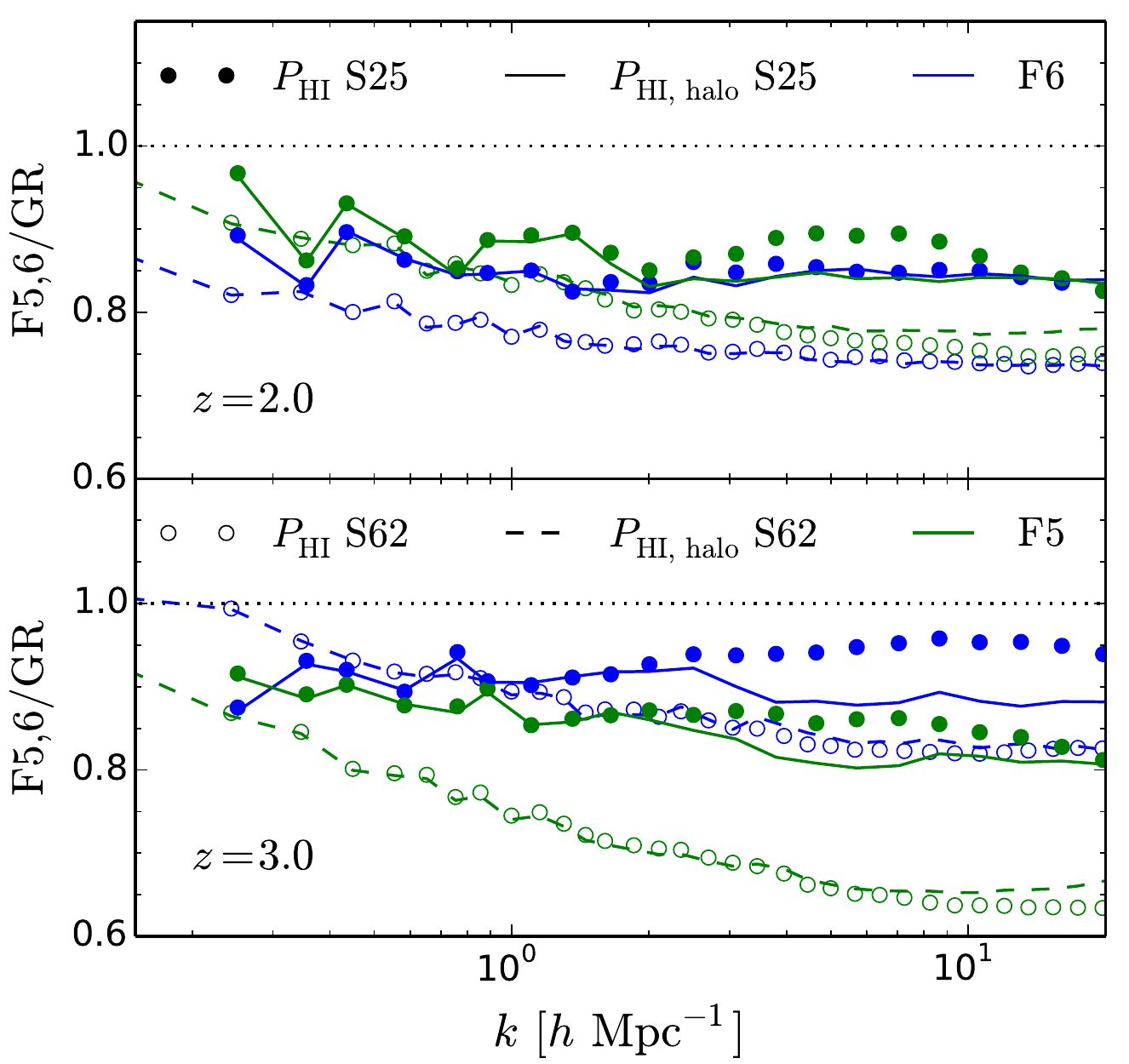}
  \caption{The ratios of the F5 (green) and F6 (blue) HI power spectra w.r.t. GR for the \textit{actual} real-space HI power spectrum (symbols) and the \textit{halo} HI power spectrum (lines), $P_{\rm HI, halo}$, as defined in the text, at $z=2.0$ (upper panel) and $z=3.0$ (lower panel). 
  Solid lines and full circles show the results from {S}25, while dashed lines and empty circles show the results from {S}62.} 
  \label{fig:Fig_HIhaloPKvsactual}  
\end{figure}

To further check the above result about the different behavior of $P_{\rm HI}$ in MG and GR, we have carried out two additional tests. 

Fig.~\ref{fig:Fig_HIhaloPKvsactual} compares the HI power spectrum (the {\it actual} $P_\mathrm{HI}$) in real space with a \textit{halo} HI power spectrum, $P_\mathrm{HI,\,halo}$, calculated by assuming that for each halo all the HI contained in it is at its center. As {shown in this figure},  the ratios of $P_\mathrm{HI,\,halo}$ for $k< 2\,h\,\mathrm{Mpc}^{-1}$ are very similar (within a few \%) to those from the {\it actual} $P_\mathrm{HI}$, confirming that the model differences in $P_{\rm HI}$ at large scales are determined by halo clustering. However, at even larger $k$, the $P_\mathrm{HI,\,halo}$ results start to deviate from the {\it actual} $P_\mathrm{HI}$ because the former do not account for the spatial distribution of HI inside the halos (see, e.g., Ref.~\cite{Villaescusa-Navarro:2018vsg}), and this affects the relative difference between the $f(R)$ models and GR. Since the differences in the {\it actual} power spectrum among the models closely follow those in $P_\mathrm{HI,\, halo}$ at large scales, we can conclude that the HI distribution is more sensitive to the clustering of halos than to the effects of baryons at these scales. As we can see from Fig.~\ref{fig:Fig_HIhaloPKvsactual}, the above result is true for both simulation set-ups, including {S}62. Because in {S}62 we do not have accurate information on the clustering of halos with masses $<10^{10}\,\mathrm{M}_\odot$, inevitably the HI power spectrum results therein are not accurate and tend to deviate from their high-resolution counterparts. 

\begin{figure}
 \advance\leftskip-0.2cm
  \includegraphics[width=.5\textwidth]{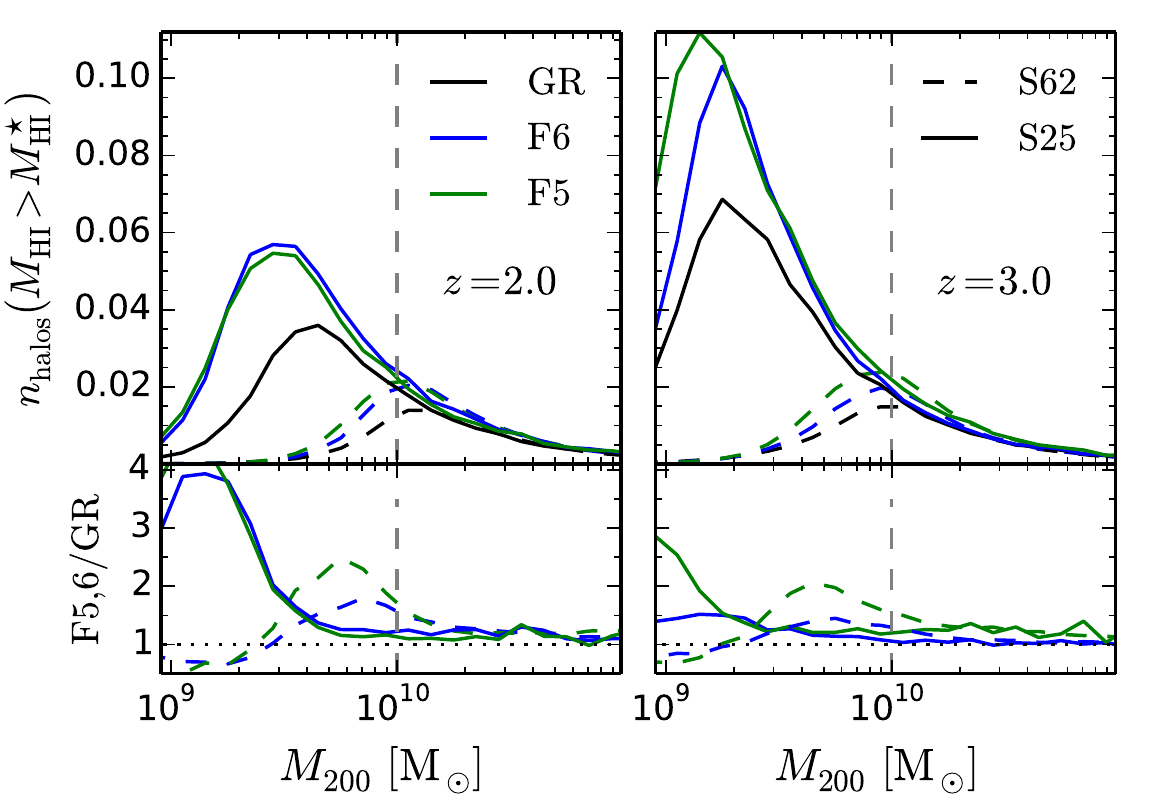}
  \caption{Number density of halos with HI mass 
  $\geq M^\star_\mathrm{HI}= 10^6\,\mathrm{M}_\odot$, for GR (black), F6 (blue) and F5 (green) at $z=2.0$ (left panel) and $3.0$ (right panel). Solid lines show the results for {S}25, while dashed lines represent those from {S}62. The vertical grey dashed lines indicate halos with $\sim 50$ particles in {S}62. We do not show the corresponding limit for {S}25 because it is at around $\sim 6\times 10^8\,\mathrm{M}_\odot$, so it is outside the halo mass range shown in the figure.}
\label{fig:Fig_number_density_HIrich}  
\end{figure}

To further show the importance of simulation resolution for accurate predictions of HI, Fig.~\ref{fig:Fig_number_density_HIrich} displays the number density of halos with HI mass $M_\mathrm{HI}\geq10^6\,\mathrm{M}_\odot$ (i.e. the halos that contribute significantly to $P_{\rm HI}$). First, we can see that in the case of {S}25 the number of HI-rich halos increases with decreasing halo mass and  peaks at a certain mass scale, $M^\mathrm{peak}_{200}$ (for GR, $M^\mathrm{peak}_{200}\sim 2\times 10^9\,\mathrm{M}_\odot$ at $z=3$ while $M^\mathrm{peak}_{200}\sim 4\times 10^9\,\mathrm{M}_\odot$ at $z=2$).  For masses lower than $M^\mathrm{peak}_{200}$, the number of HI-rich halos decreases, though the halo mass function keeps increasing. Because in S25 halos with $50$ DM particles have a mass of $\sim 6\times 10^8\,\mathrm{M}_\odot$, the halos near (including a factor of a few below) $M^\mathrm{peak}_{200}$ are well-resolved. This is not the case for {S}62, for which the peak mass $M^\mathrm{peak}_{200}$ predicted by simulations coincides with the resolution limit ($M^\mathrm{peak}_{200}\sim 10^{10}{M}_\odot$, see the vertical dashed line in Fig.~\ref{fig:Fig_number_density_HIrich}) for the simulation, implying that the decrease in the HI-rich halo abundance at $M_{200}<M^\mathrm{peak}_{200}$ is probably a resolution effect. We therefore conclude that the {S}25 simulations have a sufficient resolution for predicting the HI abundance and clustering.

Regarding the overall {low-mass} HI-rich halo abundance for {S}25, we find that at $z=3$ it is higher in F5, F6 than in GR, suggesting that HI is distributed in a less clustered way in the MG models. At $z=2$, the number of HI-rich halos in F6 is slightly larger than that in F5 for halo masses $\gtrsim 2\times10^9\,\mathrm{M}_\odot$, which may explain why the HI power spectrum is a little more suppressed in F6 than in F5 at this redshift.

Another interesting observation from Fig.~\ref{fig:Fig_number_density_HIrich} (see the results for {S}25 in the low panel) is that the HI-rich halo abundances for F6 and F5 are close to each other at $z=2$ and $z=3$. This can explain why we find a very similar degree of suppression in the power spectra of these two models in the previous subsection. This behavior of the HI-rich halo mass function can be understood as follows. Due to its less efficient chameleon screening, one may naively expect that F5 is able to turn more initial density peaks into halos (above a certain mass) than F6, and so this model always has more HI-rich halos than F6 at a given time. However, continuously decreasing the efficiency of screening cannot increase the number of low-mass halos indefinitely due to increased merger rates and because the maximum enhancement of the strength of gravity is $4/3$ in $f(R)$ gravity. Furthermore, the overall HI available to form inside halos is also limited. Fig.~\ref{fig:Fig_number_density_HIrich} implies that there is an upper limit on the number of HI-hosting halos that can be produced and making structures unscreened earlier will not help to increase the HI halo abundance beyond this `saturation point'. This is particularly evident from the results at $z=2$, where F5 and F6 have nearly identical distributions. This {result} is interesting since two models with very different screening strengths, such as F6 and F5, can end up having the same constraint.

\subsection{Observational forecast}
\label{sec:SKAres}

\begin{table}
\begin{tabular}{c c c} 
  \hline\hline
\quad\quad S/N \quad & \quad$z=3.0$\quad\quad &  \quad$z=2.0$\quad\quad \\ 
 \hline
  F6 & 4.3 & 10.1  \\ 
  \hline
 F5 & 5.5 & 9.9   \\ 
  \hline\hline
\end{tabular}
\caption{The integrated S/N ratios for distinguishing a MG model from GR using redshift space $P_{\rm HI}(k)$ with $k_\mathrm{max} = 2\,h\,\mathrm{Mpc}^{-1}$, with $({\rm S/N})^2\equiv\sum^{k_\mathrm{max}}_{k_\mathrm{min}} \left[P^\mathrm{MG}_\mathrm{HI}(k)-P^\mathrm{GR}_\mathrm{HI}(k)\right]^{2}/\sigma^2(k)$; here $\sigma(k)$ is the expected $1\sigma$ error from {\sc ska}1-{\sc mid} for $1000$ hours, while $k_\mathrm{min}$ is set by the value of the box length $L=25\,h^{-1}\,\mathrm{Mpc}$ of our high-resolution simulations, $k_\mathrm{min}=2\,\pi/L$. The above results are calculated using the redshift space HI power spectra measured from {S}25 only.}
\label{table:SKA_S/N}
\end{table}

To understand the extent to which future measurements of $P_\mathrm{HI}$ in redshift space can distinguish F6 and F5 from GR, we estimate the $1\sigma$ errors on the power spectrum expected from the instrumental noise of {\sc ska}1-{\sc mid} radio telescope \cite{Paper-SKA1-MID} for GR, following the method given in \cite{Carucci:2016yzq,21cm-non-st-phys_1, 21cm-non-st-phys_2, 21cm-non-st-phys_3} and using a realistic baseline densities computed in \cite{Bull:2014rha}. We compute the expected $1\sigma$ errors for  $1000$ observing hours, as shown by the shaded areas in the central lower subpanels of Fig.~\ref{fig:Fig_HIPK_HMF}. In the calculation we have used the monopole of the HI redshift space power spectrum for GR measured from the S25 simulation. Comparing the errors with the $P_\mathrm{HI}$ ratios w.r.t. GR in redshift space {for S25} (solid lines), we find that F6 and F5 can be both distinguished from GR at $z=2$ and for $k<2\,h\,\mathrm{Mpc}^{-1}$ by using a $1000$-hour integration. The integrated signal-to-noise (S/N) ratios for distinguishing MG to GR are shown in Table~\ref{table:SKA_S/N} for the two redshifts considered in this analysis.

\subsection{Uncertainties in subgrid physics}
\label{sec:subgrid_approximation}

Current state-of-the-art simulations are still unable to fully resolve the formation and evolution of stars and galaxies from first principles, largely due to the huge dynamical range between the scales at which star formation and black hole accretion take place and the scales which must be covered in order to realistically reproduce the large-scale structure of the Universe. As a result, these simulations rely on simplified models to approximate the main small-scale baryonic processes, such as stellar and black hole evolution and their feedback, beyond the resolution limits. These are referred to as `subgrid' recipes -- such as the IllustrisTNG galaxy formation model used in this work -- and usually include a number of free parameters that are calibrated against a set of observational data.

The IllustrisTNG model was calibrated against the stellar mass function, stellar mass fraction, galaxy central black hole masses and gas fraction, galaxy sizes at redshift $z=0$, and the cosmic star formation rate history. Not surprisingly, the approximate nature of the subgrid physics can introduce uncertainties and degeneracies in the results. Two of the questions, which follow immediately, are thus: (1) can the IllustrisTNG subgrid model, which is calibrated for $\Lambda$CDM, be used in the $f(R)$ simulations without any changes, and (2) does the simplified implementation of baryonic physics severely affect the signal expected from modified gravity? The first question was addressed in Ref.~\cite{arnold2019}, where it was found that a re-calibration is not necessary, because of the large uncertainties in observational data and the relatively modest effect of modified gravity on the observables used for calibration. Concerning the second question, we do expect the HI distribution to depend on the underlying galaxy formation model. Unfortunately, there is no easy way to quantify the effect of changing the model without running a large number of simulations with different subgrid physics. This latter effort is beyond the scope of this work and would still leave plenty room for uncertainties. We will therefore limit our discussion below to qualitative comments on the possible effects of changing subgrid physics on our results.

An important point to understand is to what extent the uncertainties in AGN and stellar feedback can affect the results above. We consider first AGN feedback, which can expel HI from the centers of massive halos or ionize the gas \cite{Villaescusa-Navarro:2018vsg}, hence affecting its clustering. While the impact of AGN feedback on the total matter power spectrum $P_\mathrm{tot}(k)$ at $z=0$ is found to be strong in previous works (see, e.g., \cite{2011MNRAS.415.3649V}), more recent high resolution simulations of galaxy formation, such as {\sc eagle} (see Fig.~1 in \cite{Hellwing:2016ucy}) and IllustrisTNG (see Fig.~7 in \cite{springel2018}), agree remarkably well and find it to be less than $1\%$ on scales $k\lesssim 2\,h\,\mathrm{Mpc}^{-1}$. At higher redshifts, which we are interested in here, the effect is even weaker.

Moreover, at scales $k<2\,h\,\mathrm{Mpc}^{-1}$ (cf.~Section~\ref{sec:Addtests}) the MG effects on the HI clustering are mainly due to changes in low-mass ($<10^{10}\,\mathrm{M}_\odot$) unscreened objects which do not host AGN. More massive, AGN-hosting halos tend to be screened, particularly at high redshift. The halo mass at which chameleon screening becomes efficient can be estimated using the fitting formula presented in Ref.~\cite{Mitchell:2018qrg}, and we find that halos more massive than $2\times 10^{9}\,\mathrm{M}_\odot$ and $2.5 \times 10^{10}\,\mathrm{M}_\odot$ in F6 are screened at $z=3$ and $z=2$, respectively, while AGN feedback mainly influence halos of $M_{200} > 4\times 10^{12}\,\mathrm{M}_\odot$ \cite{Villaescusa-Navarro:2018vsg}. As a result, AGN feedback and modified gravity affect two different regimes of halo masses.

We do not expect moderate changes in stellar feedback to have a strong impact on the relative difference between the HI power spectra of the models either. This is because at large scales ($k<2\,h\,\mathrm{Mpc}^{-1}$) the clustering of HI closely mimics that of the dark matter halos which host HI, while stellar feedback mainly affects the distribution of gas and stars within the halos and thus smaller scales.
Therefore, at least for the feedback prescription used in IllustrisTNG, stellar feedback does not seem to significantly change the key signature of modified gravity on dark (and total) matter clustering. 
Much stronger variations to the feedback mechanism may alter the HI content within halos of a given mass range and consequently the HI power spectrum; such strong changes would nevertheless lead to tensions with the low redshift observables used to tune the IllustrisTNG model \cite{pillepich2018}. 
We also note that in Ref.~\cite{Altay:2013tda} the authors analyzed the impact of different physical processes on the HI column density, and found that a significant effect (note that the variations of the subgrid physics parameters in that work are extreme) is only observed in the regime of very high column density, corresponding to the very inner regions of massive halos where the AGN feedback is strong.

From the discussion above, we expect the main conclusions of this work to be relatively robust against changes of subgrid physics parameter in the simulations. This said, given that HI clustering is a promising probe of cosmology for the future, it would be useful to study in greater detail how it is affected by the uncertain subgrid physics associated with galaxy formation -- not only in the context of modified gravity, but also for the $\Lambda$CDM model itself.

\section{Summary and conclusions}
\label{sec:SummConc}

The screening mechanism of chameleon-type MG models, such as $f(R)$ gravity, is particularly efficient at high redshift and for massive objects. Consequently, the cosmological probes proposed to date focus primarily on low redshift. The constraining power, mainly from high-mass objects, is nevertheless still limited due to the efficient screening in these objects, while it is difficult to obtain accurate cosmological data for less-screened, low-mass objects.

In this paper, we explore a different approach to constrain $f(R)$ gravity by making use of the fact that the low-mass end of the halo mass function is enhanced in this MG model already at redshift $z\sim2 - 3$. We propose that this enhancement should be observable through 21 cm intensity mapping and use the {\sc shybone} simulations, a set of state-of-the-art hydrodynamical simulations employing the IllustrisTNG model which were carried out for two different $f(R)$ gravity models (F5 and F6), to analyze the viability of this approach. Our results can be summarized as follows.

\begin{itemize}
\item The ratios of the overall HI abundance w.r.t. GR follow a similar trend for F6 and F5. Both models predict similar $\Omega_\mathrm{HI}$ as GR at $z\gtrsim4$. At lower redshifts, the ratios increase, reaching a maximum enhancement at $z=2$ and $z=1$ for F6 and F5 respectively before starting to decrease at even lower redshifts. The results from {S}62 are limited by the resolution, since these simulations do not resolve halos with masses $M_{200}\lesssim10^{10}\,\mathrm{M}_\odot$, which contain non-negligible amounts of HI. 

\item The HI mass enclosed in halos for GR, F6 and F5 can be  well described by the fitting formula proposed in \cite{Villaescusa-Navarro:2018vsg}, which  approaches a power law at high halo mass, and is exponentially suppressed at low masses. We provide the best-fit values of the parameters of this formula for all the models considered here at $z=2$ and $3$. Our fitting curves for GR are in agreement with those found in \cite{Villaescusa-Navarro:2018vsg}.  These results  can be used to model the HI distribution and clustering in $f(R)$ gravity using DMO simulations by painting the HI onto the center of each dark matter halo. 

\item The HI power spectrum (both in real and redshift space) in F5 and F6 is suppressed w.r.t. that measured for GR. This suppression can be detected in future 21 cm experiments such as {\sc ska}1-{\sc mid} with $1000$ hours of exposure time at a S/N ratio of $\sim 10$ at $z=2$ for both F5 and F6.   

\item The differences in the HI power spectrum closely reflect the differences in the HI-hosting halo abundance among the models. The halo mass functions in F5 and F6 are enhanced w.r.t. GR at $M_{200}\lesssim10^{12}\,\mathrm{M}_\odot$, showing that $f(R)$ gravity is able to turn more low-density peaks into halos which host HI. Since low density peaks are less clustered, HI power spectra have lower amplitudes in MG compared to GR.
The above statement is corroborated by the analysis of the HI-hosting halo power spectrum, i.e., the power spectrum calculated assuming the HI in each halo is concentrated at the center of that halo. This test shows that the HI power spectrum for $k<2\,h\,\mathrm{Mpc}^{-1}$ is influenced more by halo distribution and clustering rather than by baryonic effects.

\item  The predicted HI distribution strongly depends on the resolution of the simulations. Indeed, as can be seen from the abundance of HI-rich halos (Fig.~\ref{fig:Fig_number_density_HIrich}), the {S}62 simulations are not able to resolve halos with $M_{200}\lesssim10^{10}\,\mathrm{M}_\odot$; consequently we have no information on the HI inside such halos and these simulations are not able to give accurate predictions of the relative differences in the HI power spectra for F6 and F5 w.r.t. GR. The {S}25 simulations are, in contrast, able to resolve halos with  $M_{200}\gtrsim10^{9}\,\mathrm{M}_\odot$, predicting more HI-hosting halos than in the case of their low-resolution counterparts.
\end{itemize}

Our results indicate that 21 cm intensity mapping can prove useful in constraining $f(R)$ gravity models. For F6, for example, future 21 cm intensity experiments can offer a strong test compared with the other cosmological probes which focus on low-redshifts and massive objects.
While this seems promising, however, we note that the HI distribution may depend on the underlying galaxy formation model chosen for the simulations. In Section~\ref{sec:subgrid_approximation}, we discuss the main processes (AGN and stellar feedback) which can affect the HI clustering. However, a comprehensive study about how the different baryonic effects can affect the HI in galaxies does not exist yet, and would be an interesting topic for future work.

As a side remark, we note that 21 cm intensity mapping can also give accurate information of the expansion history \cite{21cm_wp,background_constr_1, background_constr_2}, which may be used to break potential degeneracy between modified expansion history and structure growth. This is because viable $f(R)$ models with a working chameleon screening mechanism to restore GR in the Solar system must have practically identical expansion history to $\Lambda$CDM, but structures in these models still grow differently at high $z$.

To conclude, 21 cm intensity mapping represents a new and potentially useful tool to constrain non-standard cosmological models that modify the matter and halo distribution from predictions by $\Lambda$CDM. The results of this work suggest that the HI distribution is appreciably affected by MG effects and can be used to shed light on the nature of gravity itself.

\

\textbf{Acknowledgements} We thank Carlton Baugh, Jianhua He, Tom Theuns, and Francisco Villaescusa-Navarro for useful discussions. The authors are supported by the European Research Council via Grant No. ERC-StG-716532-PUNCA. B.L. is additionally supported by Science and Technology Facilities Council (STFC) Consolidated Grant No. ST/P000541/1. This work used the DiRAC@Durham facility managed by the Institute for Computational Cosmology on behalf of the STFC Distributed Research utilizing Advanced Computing (DiRAC) High Performance Computing (HPC)  Facility (\url{www.dirac.ac.uk}). The equipment was funded by Department for Business, Energy and Industrial Strategy (BEIS) capital funding via STFC capital Grants No. ST/K00042X/1, ST/P002293/1, ST/R002371/1, and ST/S002502/1, Durham University and STFC operations Grant No. ST/R000832/1. DiRAC is part of the National e-Infrastructure.


\begin{thebibliography}{20}
\bibitem{will2004} 
  C.~M.~Will,
  Living Rev.\ Rel.\  {\bf 17}, 4 (2014).
  
\bibitem{koyama_MG_review} 
  K.~Koyama,
  Rept.~Prog.~Phys.~{\bf79}, 046902 (2016).
\bibitem{cluster_constraints_1} 
  M. Cataneo {\it et al.}, Phys.~Rev.~D{\bf92}, 044009 (2015). 
\bibitem{cluster_constraints_2}
H.~Wilcox~{\it et al.}, Mon.~Not.~R.~Astron.~Soc., {\bf452}, 1171 (2015).
\bibitem{cluster_constraints_3} X.~Liu~{\it et al.}, Phys.~Rev.~Lett., {\bf117}, 051101 (2016).
\bibitem{cluster_constraints_4} S.~Peirone {\it et al.}, Phys.~Rev.~D{\bf 95}, 023521 (2017).  
  
\bibitem{low_density_constraints_1}
  L.~Lombriser~{\it et al.}, Phys.~Rev.~Lett.,{\bf114}, 251101 (2015).
\bibitem{low_density_constraints_2} Y.~Cai {\it et al.}, Mon.~Not.~R.~Astron.~Soc., {\bf451}, 1036 (2015). 
\bibitem{low_density_constraints_3} M.~Cautun~{\it et al.}, Mon.~Not.~R.~Astron.~Soc., {\bf476}, 3195 (2018).
\bibitem{low_density_constraints_4} J.~Armijo~{\it et al.}, Mon.~Not.~R.~Astron.~Soc., {\bf478}, 3627 (2018). 
\bibitem{low_density_constraints_5} C.~Hernandez-Aguayo~{\it et al.}, Mon.~Not.~R.~Astron.~Soc., {\bf479}, 4824 (2018).
  
\bibitem{terukina2014}%
  \BibitemOpen
  \bibfield  {author} {A.}~{{Terukina}} {\it et al.}, {JCAP} \textbf {\bibinfo {volume} {4}},\ \bibinfo {eid} {013} (\bibinfo {year} {2014}).
  
\bibitem{buchdahl1970} H.~A.~Buchdahl, Mon.\ Not.\ Roy.\ Astron.\ Soc.\ {\bf150}, 1 (1970).

\bibitem{thin_shell_screening} P.~Brax {\it et al.}, Phys.~Rev.~D{\bf86}, 044015 (2012).

\bibitem{Hu:2007nk} 
  W.~Hu and I.~Sawicki, Phys.~Rev.~D{\bf 76}, 064004 (2007).
\bibitem{chameleon_1} 
  J.~Khoury and A.~Weltman,
  Phys.\ Rev.\ D {\bf 69}, 044026 (2004).
\bibitem{chameleon_2} J.~Khoury and A.~Weltman, Phys.Rev.Lett., {\bf 93}, 171104 (2004).
\bibitem{schmidt2010} {F.}~{{Schmidt}}, \prd \textbf {\bibinfo {volume} {81}},\
  \bibinfo {eid} {103002} (\bibinfo {year} {2010}).
\bibitem{zhao2011} {G.-B.}~{{Zhao}} {\it et al.}, {Phys. Rev. Lett.}\ \textbf {\bibinfo {volume} {107}},\ \bibinfo {eid} {071303} (\bibinfo {year} {2011}).
\bibitem{li2011}{Y.}~{{Li}}\ and\ {W.}~{{Hu}}, \prd \textbf {\bibinfo {volume} {84}},\ \bibinfo {eid} {084033} (\bibinfo {year} {2011})
\bibitem{lombriser2013} {{L.}~{{Lombriser}}} {\it et al.}, \prd \textbf {\bibinfo {volume} {87}},\ \bibinfo {eid} {123511} (\bibinfo {year} {2013}).
\bibitem{puchwein2013}{{E.}~{{Puchwein}}} {\it et al.}, \mnras\textbf {\bibinfo {volume} {436}},\ \bibinfo {pages} {348} (\bibinfo {year} {2013}).
\bibitem{arnold2014}{{C.}~{{Arnold}}} {\it et al.}, \mnras \textbf {\bibinfo {volume} {440}},\ \bibinfo
  {pages} {833} (\bibinfo {year} {2014}).
\bibitem{arnold2018}{{C.}~{{Arnold}}} {\it et al.}, \mnras \textbf {\bibinfo {volume} {483}},\ \bibinfo
  {pages} {790} (\bibinfo {year} {2019}).
\bibitem{Paper-SKA1-MID} P.~E.~Dewdney {\it et al.} (2013), SKA Organisation, \href{http://skatelescope.org}{http://skatelescope.org}.   
\bibitem{Santos:2017qgq} 
  M.~G.~Santos {\it et al.} [MeerKLASS Collaboration],
  arXiv:1709.06099 [astro-ph.CO].  
\bibitem{2013A&A...556A...2V} M. P. van Haarlem {\it et al.}, A\&A {\bf 556}, A2 (2013).
\bibitem{2014SPIE.9145E..22B} K. Bandura {\it et al.}, {Society of Photo-Optical Instrumentation Engineers (SPIE) Conference Series} {\bf 9145}, {914522} (2014).

\bibitem{2013MNRAS.434.1239B} R.~A. Battye {\it et al.}, Mon.\ Not.\ Roy.\ Astron.\ Soc.\ {\bf 434}, no. 2, 1239-1256 (2013).

\bibitem{Bharadwaj:2000av} 
  S.~Bharadwaj {\it et al.},
  J.\ Astrophys.\ Astron.\  {\bf 22}, 21 (2001).

\bibitem{Loeb:2008hg} 
  A.~Loeb and J.~S.~Wyithe,
  Phys.\ Rev.\ Lett.\  {\bf 100}, 161301 (2008).  

\bibitem{Bull:2014rha} 
  P.~Bull {\it et al.},
  Astrophys.\ J.\  {\bf 803}, no. 1, 21 (2015).    
\bibitem{Santos:2015gra} 
  M.~G.~Santos {\it et al.},
  PoS AASKA14 019 (2015).  

\bibitem{Villaescusa-Navarro:2018vsg} 
  F.~Villaescusa-Navarro {\it et al.},
  Astrophys.\ J.\  {\bf 866}, no. 2, 135 (2018). 
\bibitem{21cm-non-st-phys_1}  F.~Villaescusa-Navarro {\it et al.},
  Astrophys.\ J.\  {\bf 814}, no. 2, 146 (2015).
  
\bibitem{21cm-non-st-phys_2}  I.~P.~Carucci {\it et al.},
  JCAP {\bf 1507}, no. 07, 047 (2015).
\bibitem{21cm-non-st-phys_3} I.~P.~Carucci {\it et al.},
  JCAP {\bf 1712}, no. 12, 018 (2017).  
\bibitem{MF-non-stand-phys_1} 
  E.~McDonough and R.~H.~Brandenberger,
  JCAP {\bf 1302}, 045 (2013).   
\bibitem{MF-non-stand-phys_2} Y.~Wang {\it et al.},
  PoS AASKA {\bf 14}, 033 (2015).

\bibitem{Brax:2008hh} 
  P.~Brax {\it et al.},
  Phys.\ Rev.\ D {\bf 78}, 104021 (2008).
\bibitem{Wang:2012kj} 
  J.~Wang {\it et al.},
  Phys.\ Rev.\ Lett.\  {\bf 109}, 241301 (2012).
\bibitem{Ceron-Hurtado:2016jrp} 
  J.~J.~Ceron-Hurtado {\it et al.},
  Phys.\ Rev.\ D {\bf 94}, no. 6, 064052 (2016).
\bibitem{weak-field_approx} I.~Sawicki and E.~Bellini, Phys. Rev. D {\bf 92}, 084061 (2015). 
\bibitem{GW} 
  B.~P.~Abbott {\it et al.} [LIGO Scientific and Virgo and Fermi-GBM and INTEGRAL Collaborations],
  Astrophys.\ J.\  {\bf 848}, no. 2, L13 (2017). 
\bibitem{Burrage:2017qrf} 
  C.~Burrage and J.~Sakstein,
  Living Rev.\ Rel.\  {\bf 21}, no. 1, 1 (2018).
\bibitem{arnold2019} C.~Arnold, M.~Leo and B.~Li, (2019). Nature Astronomy, published online, DOI:\href{https://www.nature.com/articles/s41550-019-0823-y}{10.1038/s41550-019-0823-y}. \href{https://arxiv.org/abs/1907.02977}{arXiv:1907.02977} [astro-ph.CO]

\bibitem{2010MNRAS.401..791S} V.~Springel, Mon.\ Not.\ Roy.\ Astron.\ Soc.\  {\bf 401}, 791 (2010).
 \bibitem{planck2016} P.~A.~R.~Ade {\it et al.} [Planck Collaboration],
  Astron.\ Astrophys.\  {\bf 594}, A13 (2016). 
  
\bibitem{springel2018}  V.~Springel {\it et al.},
  Mon.\ Not.\ Roy.\ Astron.\ Soc.\  {\bf 475}, 676 (2018).                                                                                                          
  
\bibitem{genel2018}  S.~Genel,  Mon.\ Not.\ Roy.\ Astron.\ Soc.\  {\bf 474}, 3976 (2018).                                                                       
\bibitem {Vogelsberger2014}     M.~Vogelsberger {\it et al.},
  Nature {\bf 509}, 177 (2014).
  
  \bibitem{vogelsberger2014b}  M.~Vogelsberger {\it et al.},
  Mon.\ Not.\ Roy.\ Astron.\ Soc.\  {\bf 444}, no. 2, 1518 (2014).                                                                                                
  \bibitem{pillepich2018} A.~Pillepich {\it et al.},
  Mon.\ Not.\ Roy.\ Astron.\ Soc.\  {\bf 475}, 648 (2018).                                                                                                   
  \bibitem{marinacci2018}  F.~Marinacci {\it et al.},
  Mon.\ Not.\ Roy.\ Astron.\ Soc.\  {\bf 480}, 5113 (2018).                                                                  
  \bibitem{nelson2018}%
D.~Nelson {\it et al.},
  Mon.\ Not.\ Roy.\ Astron.\ Soc.\  {\bf 475}, 624 (2018).  
  
  \bibitem{pillepich2018b}%
 A.~Pillepich {\it et al.},
  Mon.\ Not.\ Roy.\ Astron.\ Soc.\  {\bf 473}, no. 3, 4077 (2018).
  
  
 \bibitem{weinberger2017} R.~Weinberger {\it et al.},   Mon.\ Not.\ R.\ Astron.\ Soc.\ 465, 3291 (2017).
  

\bibitem{2013MNRAS.430.2427R} A.~Rahmati {\it et al.},  Mon.\ Not.\ Roy.\ Astron.\ Soc.\  {\bf 430}, 2427 (2013).


\bibitem{Rao:2005ab} 
  S.~M.~Rao, D.~A.~Turnshek and D.~B.~Nestor,
  Astrophys.\ J.\  {\bf 636}, 610 (2006).  
\bibitem{Lah:2007nk} 
  P.~Lah {\it et al.},
  Mon.~Not.~R.~Astron.~Soc.,  {\bf 376}, 1357 (2007).  

\bibitem{2010ApJ...721.1448S} A. Songaila and L. L. Cowie, APJ {\bf 721},1448-1466 (2010).

\bibitem{2012A&A...547L...1N} P. Noterdaeme {\it et al.}, 
A\&A, {\bf 547}, L1 (2012).  
\bibitem{Crighton:2015pza} 
  N.~H.~M.~Crighton {\it et al.},
  Mon.\ Not.\ Roy.\ Astron.\ Soc.\  {\bf 452}, no. 1, 217 (2015).
\bibitem{Villaescusa-Navarro:2014cma} 
  F.~Villaescusa-Navarro {\it et al.},
  JCAP {\bf 1409}, no. 09, 050 (2014).
\bibitem{Springel:2000qu} 
  V.~Springel, {\it et al.},
  Mon.\ Not.\ Roy.\ Astron.\ Soc.\  {\bf 328}, 726 (2001).
\bibitem{Sheth:1999su} 
  R.~K.~Sheth, {\it et al.},
  Mon.\ Not.\ Roy.\ Astron.\ Soc.\  {\bf 323}, 1 (2001).

\bibitem{He:2015bua} 
  J.~h.~He, {\it et al.},
  Phys.\ Rev.\ D {\bf 92}, no. 10, 103508 (2015).
\bibitem{Jing:2004fq} 
  Y.~P.~Jing,
  Astrophys.\ J.\  {\bf 620}, 559 (2005).  
\bibitem{arnold2019b} C.~Arnold and B.~Li (2019), \href{https://arxiv.org/abs/1907.02980}{arXiv:1907.02980} [astro-ph.CO]. 
\bibitem{Arnold:2014qha} 
  C.~Arnold, {\it et al.},
  Mon.\ Not.\ Roy.\ Astron.\ Soc.\  {\bf 448}, no. 3, 2275 (2015).
\bibitem{Carucci:2016yzq} 
  I.~P.~Carucci {\it et al.}
  JCAP {\bf 1704}, no. 04, 001 (2017).
\bibitem{2011MNRAS.415.3649V} M.~P.~van Daalen {\it et al.}, 
Mon.\ Not.\ Roy.\ Astron.\ Soc.\, {\bf 415}, 3649 (2011).
\bibitem{Hellwing:2016ucy} 
  W.~A.~Hellwing {\it et al.},
  Mon.\ Not.\ Roy.\ Astron.\ Soc.\  {\bf 461}, L11 (2016).
\bibitem{Mitchell:2018qrg} 
  M.~A.~Mitchell {\it et al.},
  Mon.\ Not.\ Roy.\ Astron.\ Soc.\  {\bf 477}, no. 1, 1133 (2018).

\bibitem{Altay:2013tda} 
  G.~Altay {\it et al.},
  Mon.\ Not.\ Roy.\ Astron.\ Soc.\  {\bf 436}, 2689 (2013).
  
\bibitem{21cm_wp}
  R.~Ansari~{\it et al.} (2018), arXiv: 1810.09572.

\bibitem{background_constr_1} 
  F.~B.~Abdalla {\it et al.} [Cosmology SWG Collaboration],
  arXiv:1501.04035. 
  
\bibitem{background_constr_2} P.~Bull,
  Astrophys.\ J.\  {\bf 817}, no. 1, 26 (2016).  
  
\end{thebibliography}
\end{document}